\pdfoutput=1
\documentclass[11pt]{article}

\usepackage[margin=1in]{geometry}
\usepackage{amsmath,amssymb}
\usepackage{array}
\usepackage{booktabs}
\usepackage{cite}
\usepackage{graphicx}
\usepackage{float}
\usepackage{xcolor}
\usepackage{tikz}
\usepackage{pgfplots}
\pgfplotsset{compat=1.17}
\usetikzlibrary{arrows.meta,positioning,fit,backgrounds,patterns,calc}
\usepgfplotslibrary{groupplots}
\usepackage[hidelinks]{hyperref}

\newcommand{\dtask}{D_{\mathrm{task}}}
\newcommand{\dfid}{D_{\mathrm{fid}}}
\newcommand{\pfail}{p_{\mathrm{fail}}}
\newcommand{\Cstar}{C^{*}_{\tau}(\epsilon,\rho)}
\newcommand{\bestfound}{\emph{best-found certified frontier}}

\title{\textbf{ApproxBrain: Certified Sufficiency Frontiers\\ in Connectome-Constrained Neural Systems}}

\author{%
  Ali Mahdavi \\
  Department of Computer Engineering\\
  SRC, Islamic Azad University\\
  Tehran, Iran \\
  \texttt{ali.mahdavi@iau.ir} \\
  \and
  Azadeh Zamanifar \\
  Department of Computer Engineering\\
  SRC, Islamic Azad University\\
  Tehran, Iran \\
  \texttt{azamanifar@iau.ac.ir} \\
  \and
  Amir Farhad Farhadi \\
  School of Computer Engineering \\
  Iran University of Science and Technology \\
  Tehran, Iran \\
  \texttt{amfarhadi@mail.iust.ac.ir} \\
  \and
  Omid Kashefi \\
  Meta \\
  CA, USA \\
  \texttt{kashefi@meta.com} \\%
}
\date{\today}

\begin{document}
\maketitle

\begin{abstract}
\textbf{Background.} A connectome specifies the substrate available to a nervous system, not how much
of it a given computation requires. Reported compact sufficient circuits are point estimates: one
subset, one task, one fidelity threshold, one parameterisation.

\textbf{Methods.} We formalise computational sufficiency as a cost-indexed family of neuron subsets
conditioned on a task and a tolerance pair (permitted degradation $\epsilon$, permitted failure
fraction $\rho$), separate task adequacy from phenotype fidelity, and certify candidates on held-out
parameter draws with one-sided bounds rather than reporting a single minimum. We measure three
quantities a single circuit cannot express: non-decomposability, degeneracy, and an
intervention-derived role taxonomy.

\textbf{Results.} On a released walking simulator, the published $148$-neuron circuit is certified
sufficient with margin (mean $\dtask = 0.017$ over $32$ held-out draws, no failure), no $146$- or
$147$-neuron subset of it is sufficient, and removing any one of its three interneurons destroys
adequacy.
Complement splitting recovers the published set exactly in three of seven seeds and reaches its size
in five, within $66$--$97$ evaluations, against $221$ neurons for the bundle hierarchy at the same
budget. Varying only the task changes which neurons are needed: our best-found solutions retain $155$
and $149$ neurons and share only $3$ of $11$ searchable neurons. Relaxing the tolerance reveals
alternative sufficient circuits, and the whole-CNS interface alone fails every replicate while $93$
search-chosen neurons give a certified solution.

\textbf{Conclusions.} Sufficiency is set-valued and conditioned jointly on task and tolerance, and no
single search strategy dominates across regimes. A frontier is interpretable only alongside its
interface-only baseline, its tolerance and its budget. Every result here is model-relative.
\end{abstract}

\section*{Contributions and evidence}
\addcontentsline{toc}{section}{Contributions and evidence}

Table~\ref{tab:claims} states the contributions and the evidence supporting each.
Tables~\ref{tab:results-ledger} and~\ref{tab:results-ledger2} list the individual results. Every entry
is a property of a simulator under sampled parameters, not a measurement of an animal.

\begin{table}[h]
\centering
\footnotesize
\begin{tabular}{@{}lp{5.5cm}ll@{}}
\toprule
ID & Contribution & Evidence & Scope \\
\midrule
C1 & Sufficiency as a certified, phenotype-conditioned, failure-constrained family & Formalism & formal \\
C2 & Greedy is exact on tractable systems; the hierarchy is not & Exhaustive enumeration, $3$ seeds & synthetic; reversal \\
C3 & Non-decomposability explains why neuron ranking fails & Signed interaction decomposition & synthetic \\
C4 & Sufficient subsets are multiple rather than unique & Enumerated families; walking circuit at $\epsilon=0.10$ & model-relative \\
C5 & Necessity-sufficiency taxonomy separates conflated roles & Role table & synthetic \\
C6 & Walking circuit certified sufficient with margin; size-minimal in the tested deletion class; its three interneurons individually necessary, not partner-substitutable & Benchmark~0, $32$ held-out draws & model-relative \\
\bottomrule
\end{tabular}
\caption{Contributions. C6 is the circuit-level claim and rests on held-out certification of a
released simulator. C1 to C5 are what make the framework novel relative to
existing sufficiency screens.}
\label{tab:claims}
\end{table}

\begin{table}[h]
\centering
\footnotesize
\begin{tabular}{@{}lp{5.5cm}ll@{}}
\toprule
ID & Result & Evidence & Scope \\
\midrule
R1 & Scale-aware search traces a graded certified real-circuit frontier & Table~\ref{tab:real} & model-relative \\
R2 & Eleven-fold more substrate removed than budgeted greedy & Matched budget & model-relative \\
R3 & Search indistinguishable from random removal (all-excitatory VNC model) & Table~\ref{tab:random} & negative result \\
R4 & Real wiring no more non-decomposable than rewiring (all-excitatory) & $60$ pairs & negative result \\
R5 & On the complete male CNS, a $93$-neuron augmentation beyond the protected interface is certified within $0.47$ & Table~\ref{tab:cns} & model-relative \\
R6 & $2{,}100\times$ more substrate removed than budgeted greedy on the CNS & Matched budget & model-relative \\
R7 & Search beats random removal at the interface-failure point ($0.0069$ vs $0.727$); margin ${\sim}1.6\times$ at $157{,}577$ & Table~\ref{tab:cns-random} & model-relative \\
R8 & CNS pair-interaction ordering exceeds its rewiring control; both negligible in magnitude & $80$ pairs & model-relative \\
R9 & The CNS interface alone is insufficient ($0.746$, $\pfail=1.0$); the VNC interface alone suffices ($0.051$) & interface-only baseline & model-relative \\
R10 & $93$ search-chosen neurons beyond the interface give a certified $0.462$ & Table~\ref{tab:cns} & model-relative \\
R11 & The published $148$-neuron walking circuit is sufficient with margin ($0.017$, $\pfail=0$, $32/32$) & Table~\ref{tab:walking} & model-relative \\
R12 & No $146$- or $147$-neuron deletion subset of the published circuit suffices; $148$ smallest tested & Table~\ref{tab:walking} & model-relative \\
\bottomrule
\end{tabular}
\caption{Results: search, duplication and scale. Every row is a property of a simulator under sampled
parameters, not a measurement of an animal.}
\label{tab:results-ledger}
\end{table}

\begin{table}[h]
\centering
\footnotesize
\begin{tabular}{@{}lp{5.5cm}ll@{}}
\toprule
ID & Result & Evidence & Scope \\
\midrule
R13 & The three interneurons are not interchangeable with their same-type bilateral partners & Table~\ref{tab:walking} & negative result \\
R14 & Additions beyond the three are inert: $221$ supersets score as $148$, $221$ omissions fail & Table~\ref{tab:walking} & model-relative \\
R15 & The fourth interneuron improves fidelity ($0.414\to0.373$) without being necessary & Table~\ref{tab:walking} & model-relative \\
R16 & Complement splitting recovers the published minimum in $66$--$97$ evaluations in five of seven seeds; H-CABES stops at $221$ in $400$ & Fig.~\ref{fig:recovery} & model-relative \\
R17 & Sibling batching costs search quality: the same four seeds reach $148$ once, not three times & Section~\ref{sec:recovery} & negative result \\
R18 & Distinct $148$/$149$-neuron alternatives exist but only clear $\epsilon=0.10$, so the certified minimum is unique at $\epsilon=0.05$ & Section~\ref{sec:degeneracy} & model-relative \\
R19 & Neuron $86$ is a partial substitute for $1167$ ($0.912\to0.078$ as a replacement, $0.017\to0.0004$ as an addition) & Section~\ref{sec:degeneracy} & model-relative \\
R20 & The published circuit is near pairwise-decomposable: one dominant pair ($\gamma=-0.89$), small third-order term ($+0.058$) & Fig.~\ref{fig:interactions} & model-relative \\
R21 & $24$ random pairs give $\gamma=0.000$ exactly, so uniform pair sampling is extremely inefficient here & Fig.~\ref{fig:interactions} & negative result \\
R22 & Task conditioning on a real model: best-found solutions retain $155$ versus $149$ neurons and share only $3$ of $11$ searchable neurons & Fig.~\ref{fig:tasks} & model-relative \\
R23 & Wave-batched splitting is dominated by H-CABES on the CNS, and its only strict certificate is degenerate & Fig.~\ref{fig:cns-frontier} & negative result \\
\bottomrule
\end{tabular}
\caption{Results: circuit structure, task conditioning and whole-CNS scale. R17 and R23 concern the
search methods introduced here and are negative; they are reported for that reason.}
\label{tab:results-ledger2}
\end{table}

\section{Introduction}

\paragraph{Mapping is no longer the endpoint.}
Connectomes for the adult fly central nervous system and its ventral nerve cord are now effectively
complete at scale, and are distributed with cell-type, neurotransmitter, and neuropil annotations
\cite{scheffer2020hemibrain,dorkenwald2024flywire,schlegel2024annotation,winding2023insect,azevedo2024female,berg2026male,cheong2026manc,court2020nomenclature}. The brain has moved from an incompletely
observed graph to an object we can computationally interrogate.

\paragraph{A connectome describes possibility, not sufficiency.}
A wiring diagram states which connections exist. It does not state which subset of neurons is
sufficient to produce a given computation at a given fidelity. Sufficiency is a different and in
several respects more tractable question than necessity: it asks what can be removed while a
specified phenotype survives, rather than what must be present.

\paragraph{Recent work shows the opportunity, and its limits.}
Pugliese et al. describe their procedure as a ``computational sufficiency screen'' and prune a
connectome-constrained simulation to a minimal necessary-and-sufficient walking circuit
\cite{pugliese2026}. Li et al. fit a connectome-constrained model to brain-wide recordings and
identify the neurons and synapses that sustain spontaneous activity \cite{li2026}. Effectome analysis identifies circuits
predicted to dominate dynamics from connectome structure \cite{effectome2024}, and
connectome-constrained models now predict neural activity and sensorimotor processing from wiring
alone \cite{shiu2024model,beiran2025prediction}. Each of these is a
single point in a larger space: one circuit, one phenotype, one threshold, one parameter draw. None
reports how the required substrate varies with tolerated error, how many alternative sufficient
circuits exist, or whether the computation decomposes into independent units.

\paragraph{The gap.}
What is missing is (i) a formulation in which the fidelity-cost tradeoff is itself the object of
study, (ii) a search method sample-efficient enough to estimate that tradeoff under a bounded
simulator budget, and (iii) outputs that quantify interaction structure, multiplicity, and
functional role rather than nominating a single circuit.

\paragraph{ApproxBrain.}
We formalise sufficiency as a certified family of subsets (Section~\ref{sec:formalism}), introduce
the sufficiency frontier, define non-decomposability and degeneracy as measurable quantities, and
give an adaptive search toolkit rather than a single method: a hierarchical beam search whose
single-move restriction is exactly greedy backward elimination, and a complement-splitting group
test for the sparse critical-core regime (Section~\ref{sec:method}). We validate against exactly
enumerable frontiers, against a dynamical analogue with a hidden oscillator, and against a released
walking simulator with a published reduced circuit (Section~\ref{sec:results}). The experiments
decide between the two searches: neither dominates, and which regime favours which search is itself a
finding of this paper.

\paragraph{Standing commitment.}
Every result in this paper is relative to a specified simulator, intervention operator, phenotype,
evaluation protocol, cost, and search budget. We report sufficiency \emph{within a
connectome-constrained model}. We do not claim that a reduced simulation is what an animal needs,
that removed neurons are unnecessary, or that any discovered circuit is unique.

\medskip
\noindent\emph{A connectome tells us what a nervous system contains. ApproxBrain asks how much of it
a particular computation requires.}

\section{Related work}
\label{sec:related}

\paragraph{Connectome-constrained simulation and minimal circuits.}
Rate-based and spiking simulations constrained by measured connectivity are now used to attribute
function to identified neurons, and greedy removal under a phenotype threshold has produced compact
circuits for specific behaviours \cite{pugliese2026,shiu2024model,beiran2025prediction}. This
literature supplies our primary benchmark.
Its objective, however, is typically a single boolean threshold on one phenotype, which yields one
feasible point rather than a frontier.

\paragraph{Whole-brain modelling and sufficient substrate.}
A connectome-constrained model fitted to brain-wide recordings has been used to identify the neurons
and synapses that sustain
spontaneous dynamics \cite{li2026}. This is the closest prior work to our question, and it is why we
do not claim priority for connectome-scale sufficiency analysis.

\paragraph{Dynamical dominance and structural centrality.}
Eigenmode and influence analyses identify circuits predicted to dominate dynamics
\cite{effectome2024}, and network-control principles have been used to predict neuron
function in the \emph{C.\ elegans} connectome \cite{yan2017network}. These supply the natural contrast for
our participation and role analyses: structural or dynamical prominence need not coincide with
computational sufficiency.

\paragraph{Lesion modelling.}
The dynamical consequences of lesions in cortical networks have been studied directly, and the
literature is explicit that the measured effect of an acute manipulation is not the effect of losing
the manipulated cells \cite{otchy2015acute,wolff2018promise,sauerbrei2020cortical}. We adopt the
operational discipline of that
literature, namely that clamping, outgoing-synapse blockade, and dimensional deletion are distinct
interventions.

\paragraph{Pruning and sparsification.}
Magnitude and activity-based pruning of artificial networks \cite{han2015learning,frankle2019lottery}
is
conceptually adjacent, but its objective is task loss on data, not a graded physiological phenotype
evaluated over stochastic parameter draws. We treat those methods as ranking baselines rather than as
the same problem.

\paragraph{Rate-distortion in neural systems.}
Wiring-economy and rate-distortion arguments have been applied to neural coding and to the
organization of connectomes \cite{bullmore2012economy,chklovskii2002wiring}. Our ``rate'' is retained
computational substrate, not information transmission,
and we explicitly do not propose a new neural rate-distortion theory.

\paragraph{Circuit discovery for interpretability.}
Automated circuit discovery in artificial networks \cite{conmy2023acdc} shares the goal of isolating a
subgraph responsible for a behaviour, and shares the difficulty that individually small components
can be jointly necessary.

\section{Computational sufficiency}
\label{sec:formalism}

\subsection{The estimand is a tuple}

A sufficiency result is a property of a complete tuple, not of a brain or of a model alone:
\begin{equation}
\mathcal{M} \;=\; \bigl(\mathcal{G},\Theta,\tau,I,\mathcal{B}_{\tau},C,\dtask,\dfid,H,\mathcal{H}_{\mathrm{eval}}\bigr),
\end{equation}
where $\mathcal{G}=(V,E,W)$ is the model, $\Theta$ its parameters, $\tau$ the phenotype, $I$ the
intervention operator, $\mathcal{B}_{\tau}$ the protected interface, $C$ the cost, $\dtask$ and
$\dfid$ the metrics, $H$ the hierarchy, and $\mathcal{H}_{\mathrm{eval}}$ the evaluation protocol.

\subsection{Intervention}
\label{sec:intervention}

Let $z\in\{0,1\}^{|V|}$ be a retained-neuron mask. The intervention used throughout is
\begin{equation}
W^{(z)}_{ij} \;=\; z_i\, W_{ij}\, z_j, \qquad z_i = 1 \ \ \forall i \in \mathcal{B}_{\tau}.
\end{equation}
A removed neuron is clamped to zero and contributes no outgoing signal. This choice is fixed for all
methods, because clamping a state, blocking outgoing synapses, and deleting a state variable are
different interventions with different consequences. We report at least one contrast against
synapse blockade so that the headline does not depend on the choice.

\subsection{Protected interface}

We require $S \supseteq \mathcal{B}_{\tau}$, where $\mathcal{B}_{\tau}$ contains the stimulated input
neurons and the motor readout neurons. Without this constraint a search can achieve sufficiency by
deleting the task.

\subsection{Adequacy and fidelity are distinct}
\label{sec:adequacy}

Let $q_{\tau}(y)\in[0,1]$ be a scalar phenotype score. Adequacy is the boolean event
\begin{equation}
A_{\tau}(y) \;=\; \mathbf{1}\!\left[q_{\tau}(y)\geq q_{\min} \ \wedge\ \text{stable} \ \wedge\ \text{readout present}\right],
\end{equation}
and fidelity is a graded distance in phenotype features $\phi_{\tau}(y)$ (rhythmicity, frequency,
active readout pattern, phase relations, amplitude). Accordingly we report three quantities:
\begin{align}
\dtask(z) &= \mathbb{E}_{\xi\mid\Omega_{\tau}}\!\left[\bigl(q_{\tau}(y_{0}) - q_{\tau}(y_{z})\bigr)_{+}\right],\\
\pfail(z) &= \Pr_{\xi\mid\Omega_{\tau}}\!\left[A_{\tau}(y_{z}) = 0\right],\\
\dfid(z) &= \mathbb{E}_{\xi\mid\Omega_{\tau}}\!\left[d_{\tau}\bigl(\phi_{\tau}(y_{0}), \phi_{\tau}(y_{z})\bigr)\right].
\end{align}
The separation matters empirically: the published reduced circuit reportedly preserves rhythm while
losing stimulus-dependent frequency scaling \cite{pugliese2026}. A single omnibus distortion would
hide that, and could be tuned, deliberately or not, to favour a known answer.

\subsection{Reference-valid population}

A replicate in which the intact network does not express the phenotype cannot establish
preservation. Define
\begin{equation}
\Omega_{\tau} \;=\; \bigl\{\xi : y_{0}(\xi) \text{ is stable and } A_{\tau}(y_{0}(\xi)) = 1\bigr\},
\end{equation}
estimated from the intact network only, frozen once, and reused for every mask. All expectations are
conditional on $\Omega_{\tau}$. Unstable masked runs receive maximal loss and are never silently
discarded, and the unconditional success probability of the intact model is reported separately so
that conditioning is visible.

\subsection{The sufficiency frontier}
\label{sec:frontier}

With cost $C(z)$, the problem is
\begin{equation}
z^{\star}_{\tau,\epsilon,\rho} \;\in\; \arg\min_{z} C(z)
\quad \text{subject to} \quad
\dtask(z)\leq\epsilon, \quad \pfail(z)\leq\rho, \quad z_i = 1 \ \forall i\in\mathcal{B}_{\tau}.
\end{equation}
Varying $\epsilon$ and $\rho$ gives the \emph{computational sufficiency frontier}
\begin{equation}
\Cstar \;=\; \min_{z \in \mathcal{F}_{\epsilon,\rho}} C(z),
\qquad
\mathcal{F}_{\epsilon,\rho} = \{z : \dtask(z)\leq\epsilon,\ \pfail(z)\leq\rho\}.
\end{equation}
Because the optimiser is heuristic, we report $\Cstar$ as the \bestfound{} and never as the true
minimum. The failure constraint is not cosmetic: without it, a low mean distortion can conceal rare
catastrophic or numerically unstable runs.

\subsection{Robust sufficiency}
\label{sec:robust}

A frontier for a single parameter fit may be an artefact of that fit. We therefore define a robust
variant in which sufficiency must hold across an uncertainty set $\mathcal{U}$ over $\Theta$,
\begin{equation}
z^{\star}_{\mathrm{rob}}(\epsilon) \;=\; \arg\min_z C(z)
\quad\text{subject to}\quad
\sup_{\theta\in\mathcal{U}} \dtask(z;\theta) \leq \epsilon,
\end{equation}
and report worst-case alongside mean fidelity. This converts the strongest objection to
connectome-constrained sufficiency results, namely model dependence, into a measured result.

\subsection{Novel quantities}
\label{sec:quantities}

\paragraph{Non-decomposability.}
Treat degradation as a set function over removals. For a removal context $A$ and neurons $i,j$,
\begin{equation}
\gamma_{ij}(A) \;=\; D(A) + D(A\cup\{i,j\}) - D(A\cup\{i\}) - D(A\cup\{j\}).
\end{equation}
A positive $\gamma$ means two removals are \emph{jointly} more damaging than the sum of their
individual effects: each is dispensable alone while the pair is not. This is precisely the regime in
which single-neuron ranking fails, and it is the connectome-intervention analogue of the higher-order
interaction structure that has been characterised in complex and neural systems
\cite{battiston2021physics,rodriguez2025higher,niedostatek2025mining}. A negative $\gamma$ indicates
saturating damage, the motif that recurs in the walking circuit below. We report the
signed decomposition
\begin{equation}
\text{super-additivity} \;=\; \mathbb{E}\!\left[\max(\gamma,0)\right],
\qquad
\text{sub-additivity} \;=\; \mathbb{E}\!\left[\max(-\gamma,0)\right],
\end{equation}
and use super-additivity as the \emph{non-decomposability index}. An additive circuit scores $0$; the
sign convention is essential, since an unsigned magnitude conflates a jointly-required pair with an
individually-necessary one.

\paragraph{Degeneracy.}
Sufficient sets are generally not unique, a property that in biological systems is distinct from
redundancy and is a source of robustness and evolvability \cite{edelman2001degeneracy}. We report the
number of distinct certified subsets at each
cost level, the distribution of pairwise Jaccard overlap, and the stable core (neurons present in at
least a fraction of near-optimal solutions). This makes multiplicity a reported result rather than a
caveat.

\paragraph{Functional role taxonomy.}
Classify each neuron by its marginal necessity and its participation in positive interactions,
yielding roles that connectivity and cell type do not determine: critical; synergistic (individually
dispensable, jointly required); dispensable; inert. This operationalises a distinction that
single-ablation ranking conflates by construction.

\paragraph{Higher-order structure.}
Pairwise probes can miss interactions that appear only at order three, the dynamical counterpart of
subgraph motifs that recur far more often than chance in biological networks
\cite{milo2002network}. We therefore report an
explicit search for triples whose pairwise signals are weak while their joint effect is large, and
treat the existence of such triples as a falsifiable prediction rather than an assumption.

\section{Method: an adaptive search toolkit}
\label{sec:method}

\subsection{Approach}

ApproxBrain is a framework plus two search members that address different circuit structure.
H-CABES (Hierarchical, Confidence-Aware Beam Elimination Search) is the general-purpose member for
interaction-rich, distributed sufficiency, where no small critical core exists and the search must
trade verification depth for breadth. Adaptive complement splitting (Section~\ref{sec:split}) is the
member for sparse critical cores, where a few necessary neurons must be located among thousands of
inert ones. Section~\ref{sec:walking} reports the benchmark on which the two partition the problem.
The design of H-CABES is deliberately conservative: the method should be the simplest procedure that
handles non-monotone, interaction-dominated damage under an explicit evaluation budget.

\subsection{Hierarchy}

Levels are ordered coarse to fine, from a disjoint cell-class partition down to single neurons. A
leaf is retained only if all of its ancestors are retained. Two cautions apply. Neuropil annotations
overlap and therefore form a DAG rather than a tree, so a forced disjoint partition introduces
arbitrary structure. Coarse screening can also permanently exclude a distributed circuit. We
therefore keep full support, and we include a randomly permuted-label control so that any gain from
hierarchy can be separated from the biological prior encoded in the labels.

\subsection{Procedure}

\begin{enumerate}
\item Evaluate and cache intact-network outcomes on the search replicates; freeze
$\Omega_{\tau}$.
\item Initialise the beam with the full mask and an empty Pareto archive over
$(C, \dtask)$.
\item For each beam member, generate proposals: deletion of a single group; deletion of a small
random bundle of groups; and a restore-one/delete-one swap.
\item Evaluate proposals on paired replicates using common random numbers against the intact
network, respecting the evaluation budget.
\item Retain up to $K$ diverse feasible masks by cost, estimated feasibility, and archive crowding.
\item Descend retained groups into their children and repeat.
\item Stop at the budget, or when no proposal improves the archive.
\item Re-evaluate finalists on a disjoint, frozen test replicate set and certify.
\end{enumerate}

\subsection{Why the ablation is interpretable}

With beam width $K=1$, no bundle proposals, and no swap proposals, H-CABES reduces exactly to greedy
backward elimination. The strongest baseline is therefore a special case of the method, and the
contribution of each additional move is measurable rather than asserted. Bundles address removals
that are safe only jointly; swaps address path dependence and non-monotone damage.

\subsection{Adaptive complement splitting}
\label{sec:split}

The second member treats removal as a group-testing problem. Removing the whole searchable set is
infeasible when any necessary neuron exists, so the failed bundle is split and each part retested; a
part that can be removed is discarded in a single evaluation, and only parts that contain a necessary
neuron are subdivided further.

The cost of this scheme depends on an assumption we do not satisfy, and we state it rather than
quoting the ideal bound. If feasibility were monotone in the retained set and a failed test implied
that the part contained a necessary neuron, the recursion would visit $O(k\log n)$ parts for $k$
necessary neurons among $n$ candidates, against $O(n)$ for one-at-a-time elimination. Our feasibility
predicate is explicitly \emph{not} monotone: pruning can raise the one-sided degradation score, and
removing two individually dispensable neurons can destroy the task even though removing either alone
does not. Siblings of a failed part are therefore evaluated in one batch but their successful removals
are re-tested as a union rather than assumed, and a failed union falls back to applying them one at a
time. The method consequently inherits no worst-case bound in general; it returns a locally minimal
set that certification must confirm, and its advantage on a sparse critical core is an empirical
property of that regime. What we can report is the measured cost: $66$--$97$ evaluations to reach the
published $148$ neurons on the walking model, against $221$ neurons for the bundle hierarchy at
$400$ evaluations (Section~\ref{sec:recovery}).

\subsection{Scale}

A proposal distribution tuned for a $40$-neuron toy does not search a $2{,}488$-neuron circuit. With a
fixed bundle of three, one accepted move removes $0.1\%$ of the substrate, and a single-neuron greedy
sweep consumes an entire $O(150)$-evaluation budget inside its first step; both methods then report
``one neuron removable'' at \emph{every} tolerance. That number measures the budget, not the circuit,
and it is the failure mode we hit and corrected. Two changes make the search scale-aware, and both
default to the original behaviour so the exact-frontier results are untouched.

\emph{Bundle ladder.} Rather than one fixed bundle size, each round proposes deletions at several
scales, $s \in \{\lceil f n_{\mathrm{searchable}} \rceil\}$ for fractions $f$ spanning $\{0.5, 0.25,
0.1, 0.03, 0.01\}$, coarse to fine. The archive then observes the cost-degradation tradeoff directly
instead of a single point per move.

\emph{Budgeted probing.} Exact backward elimination costs $O(n)$ evaluations per removed neuron.
We give the baseline a bounded random probe of $p$ candidates per step, so it can also take several
steps inside a matched budget; without this the comparison is decided by the budget rather than by
the search. We report the probe size, and the exact ($p=n$) variant, for every headline comparison.

\subsection{Certification}

Search-set intervals guide allocation only and are \emph{not} inferential results, because
candidates are selected adaptively. Certification uses an independent test replicate set never used
for search, surrogate fitting, or stopping, and requires both
\begin{equation}
\widehat{\dtask}^{\,\mathrm{UB}}_{\alpha} \leq \epsilon
\qquad\text{and}\qquad
\widehat{\pfail}^{\,\mathrm{UB}}_{\alpha} \leq \rho,
\end{equation}
where the first is a one-sided bootstrap upper bound on a bounded mean \cite{efron1979bootstrap} and
the second is a one-sided
Clopper-Pearson bound on a Bernoulli rate \cite{clopper1934use}, with Holm correction across the
reported set of masks and
constraints \cite{holm1979simple}.

Two properties of this protocol matter, because ``certified'' is otherwise easy
to read as a distribution-free guarantee that it is not.

First, the bound is \emph{one-sided on the mean}, and the worst held-out draw can be much worse than
the mean. For the published walking circuit the mean degradation over $32$ held-out draws is
$0.0168$, which is inside $\epsilon = 0.05$, but the \emph{largest single draw} degrades by $0.332$,
which is not. Under a worst-case-per-draw criterion that circuit would fail. The only candidate we
tested that passes both readings is the published circuit plus interneuron $86$, whose worst draw is
$0.0038$. We therefore report the per-draw maximum alongside the mean wherever a frontier point is
used to support a claim, and we do not treat a mean bound as a per-draw guarantee.

Second, the bootstrap bound is not distribution-free. A bound that assumes nothing about the
distribution is available and much weaker: with $n$ draws in $[0,1]$, a one-sided Hoeffding bound
\cite{hoeffding1963probability} at
confidence $1-\alpha$ is $\widehat{\dtask} + \sqrt{\ln(1/\alpha)/(2n)}$, which at $n = 32$ and
$\alpha = 0.05$ adds $0.216$ to the observed mean. That value does not exclude $\epsilon = 0.05$ for
any candidate here. What the certificates establish is therefore exactly this: \emph{the
candidate passes the stated held-out protocol (mean degradation within $\epsilon$ and failure
fraction within $\rho$, at confidence level $\alpha$, on replicates never used for selection)}. They
are not finite-sample distribution-free verification, and the $\pfail$ side is the stronger of the two,
since $0$ failures in $32$ draws bounds the failure rate by $0.091$ at $95\%$ confidence without
assuming a parametric form.

\subsection{Complexity}

Let $B$ be the number of distinct masks simulated, $\bar{R}$ the mean replicates per mask, and
$T_{\mathrm{sim}}$ the cost of one replicate. The dominant cost is
\begin{equation}
O\!\left(B\,\bar{R}\,T_{\mathrm{sim}}\right).
\end{equation}
Proposal generation is $O(KQn/w)$ with bit-packed masks of machine word size $w$. This contrasts with
$O(2^{n})$ for exhaustive search, $O(n^{2})$ for all pairs, and
$O(n^{2}\bar{R}T_{\mathrm{sim}})$ for fully rescored greedy, the procedure whose single-move
restriction is the beam's degenerate case; greedy's approximation behaviour on submodular objectives
is the classical reference point for such comparisons \cite{nemhauser1978analysis}. Because simulator
cost dominates, all
comparisons in this paper are stated in \emph{mask evaluations at matched budget}, not in masks
proposed.

\subsection{Out of scope}

Synapse-level masks, adaptive or plastic re-optimisation, conditional per-input routing, multi-task
atlases, whole-brain scaling, exhaustive pair search, differentiable gates, and learned surrogates
are excluded from the method of record: none is required for the
claims in Section~\ref{sec:results}.

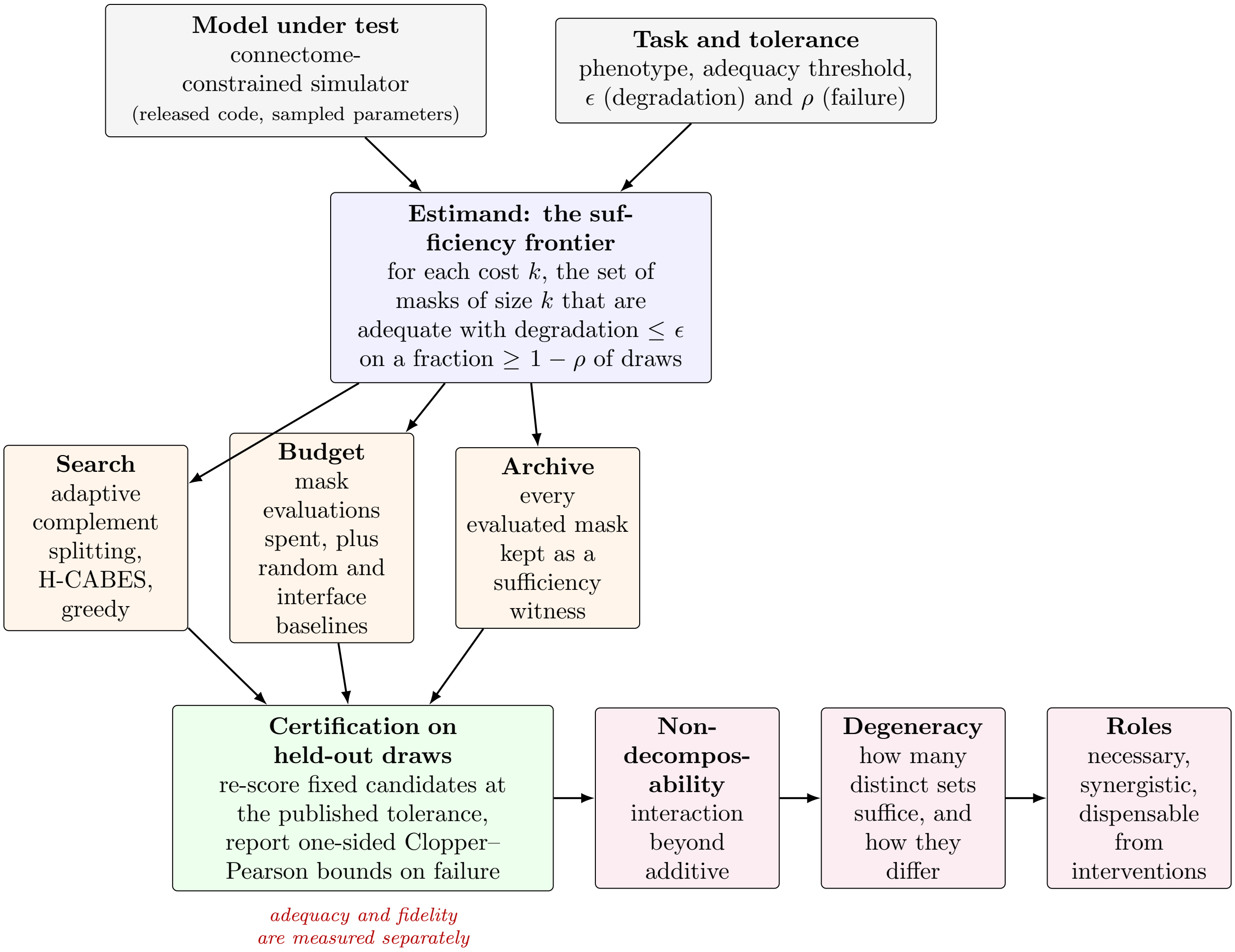
\begin{figure}[H]
\centering
\resizebox{\linewidth}{!}{%
\begin{tikzpicture}[
  font=\small,
  node distance=4mm,
  box/.style={draw, rounded corners=2pt, align=center, inner sep=4pt, minimum height=9mm,
              text width=24mm},
  wide/.style={draw, rounded corners=2pt, align=center, inner sep=5pt, minimum height=10mm,
               text width=52mm},
  lab/.style={font=\scriptsize\itshape, align=center},
  ar/.style={-{Latex[length=2mm]}, thick}
]

\node[wide, fill=black!4] (system) {\textbf{Model under test}\\connectome-constrained simulator\\\scriptsize (released code, sampled parameters)};
\node[wide, fill=black!4, right=10mm of system] (task) {\textbf{Task and tolerance}\\phenotype, adequacy threshold,\\$\epsilon$ (degradation) and $\rho$ (failure)};

\node[wide, fill=blue!6, below=9mm of $(system.south)!0.5!(task.south)$] (estimand)
  {\textbf{Estimand: the sufficiency frontier}\\for each cost $k$, the set of masks of size $k$ that are\\adequate with degradation $\le \epsilon$ on a fraction $\ge 1-\rho$ of draws};

\node[box, below=9mm of estimand, xshift=-62mm, fill=orange!8] (search)
  {\textbf{Search}\\adaptive\\complement\\splitting,\\H-CABES,\\greedy};
\node[box, right=6mm of search, fill=orange!8] (budget)
  {\textbf{Budget}\\mask\\evaluations\\spent, plus\\random and\\interface\\baselines};
\node[box, right=6mm of budget, fill=orange!8] (archive)
  {\textbf{Archive}\\every\\evaluated mask\\kept as a\\sufficiency\\witness};

\node[wide, fill=green!7, below=9mm of budget, xshift=6mm] (cert)
  {\textbf{Certification on held-out draws}\\re-score fixed candidates at the published tolerance,\\report one-sided Clopper--Pearson bounds on failure};

\node[box, right=6mm of cert, fill=purple!7] (nd)
  {\textbf{Non-}\\\textbf{decompos-}\\\textbf{ability}\\interaction\\beyond\\additive};
\node[box, right=6mm of nd, fill=purple!7] (deg)
  {\textbf{Degeneracy}\\how many\\distinct sets\\suffice, and\\how they\\differ};
\node[box, right=6mm of deg, fill=purple!7] (roles)
  {\textbf{Roles}\\necessary,\\synergistic,\\dispensable\\from\\interventions};

\draw[ar] (system) -- (estimand);
\draw[ar] (task) -- (estimand);
\draw[ar] (estimand) -- (search);
\draw[ar] (estimand) -- (budget);
\draw[ar] (estimand) -- (archive);
\draw[ar] (search) -- (cert);
\draw[ar] (budget) -- (cert);
\draw[ar] (archive) -- (cert);
\draw[ar] (cert) -- (nd);
\draw[ar] (nd) -- (deg);
\draw[ar] (deg) -- (roles);

\node[font=\scriptsize\itshape, text=red!70!black, align=center, below=1mm of cert]
  {adequacy and fidelity\\are measured separately};

\end{tikzpicture}%
}
\caption{\sloppy The framework as a pipeline. A simulator and a task with an explicit tolerance are
the inputs; the output is not one circuit but a cost-indexed family of masks, each carrying a witness
evaluation. Search and certification are separate stages: search proposes masks cheaply at a loose
tolerance, and certification re-scores a fixed candidate list at the published tolerance on parameter
draws that the search never saw. The three quantities on the right are the ones a single reported
circuit cannot express, and the last two boxes are measured separately because adequacy and fidelity
can disagree.}
\label{fig:pipeline}
\end{figure}


\section{Experimental setup}
\label{sec:setup}

\subsection{Controlled oracles with exact frontiers}

We use set functions whose optimum and interaction structure are known by construction: additive;
redundant pair (either neuron suffices, so each is individually dispensable); jointly necessary pair
(removing either alone destroys function); threshold-$k$; a small jointly-required core surrounded by
inert neurons; planted blocks; and two disjoint sufficient alternatives. For $n\leq 12$ we enumerate
all masks and compute the exact frontier, giving ground-truth regret rather than a proxy.

\subsection{Dynamical analogue}

A $40$-neuron rectified-tanh rate network contains a hidden three-neuron oscillator: an excitatory
pair with recurrent self-excitation and a slow inhibitory unit. Input and motor units are the
protected interface. Rhythmicity is measured by autocorrelation of the motor trace over a lag window.
This supplies a dynamical test in which the target circuit is unknown to the search but known to us,
and is the direct analogue of a walking central-pattern-generator benchmark.

\subsection{Real ventral nerve cord}

The real circuit is a DNg100-anchored subgraph of \texttt{manc:v1.2.1}, retrieved through neuPrint:
$2{,}488$ neurons, $299{,}069$ synapses, mean degree $120$, and a $72$-neuron protected interface
(the stimulated seeds plus $70$ motor neurons). It is simulated as sparse CSR with the same
rectified-tanh dynamics, row-normalised weights, $dt=0.1$, duration $100$, and Gaussian gain noise
$\sigma=0.05$. Sparse representation is not an optimisation here but a precondition: dense weights at
this size are $49$~MB against $9.9$~MB sparse, and the full CNS would be $220$~GB dense against
$50.6$~MB sparse. Search uses $4$ paired replicates, certification uses $40$ held-out replicates with
disjoint identifier offsets, and every cost is reported in mask evaluations.

\subsection{Complete male CNS}

The second real system is the whole \texttt{male-cns:v1.0} dataset
\cite{berg2026male,cheong2026manc}: brain and nerve cord,
$176{,}422$ neurons and $25{,}862{,}574$ connections. Its descending-neuron and motor populations
carry the anatomy established for the male nerve cord \cite{namiki2018descending,court2020nomenclature},
and its distributed-control structure has been characterised independently \cite{bates2026distributed}.
Retrieval returns one row per
\texttt{(pre, post, roi)}: a pair connected across several neuropils appears once per neuropil,
so connection weights are the \emph{sum} over those rows, not the first row. Dropping duplicates
instead of summing discards $3{,}883{,}725$ rows and undercounts about a third of the synapse weight;
after summing, the distinct-pair count matches neuPrint's own \texttt{ConnectsTo} count exactly.
Transmitter signs come from \texttt{predictedNt} (GABA and glutamate inhibitory; everything else
excitatory), giving $127{,}866$ excitatory and $48{,}556$ inhibitory neurons. The stimulated
population is the $17{,}937$ sensory neurons, the readout is the $815$ motor neurons, and those
$18{,}752$ form the protected interface. Weights are unnormalised and globally scaled by
$1/\bar d = 1/146.6$, with a tonic bias of $0.55$; $dt=0.1$, duration $30$, noise $0.05$. Mask
evaluation is forked across processes: candidates in a round are independent given the frozen
cohort, and fork shares the $\sim1$~GB CSR matrix copy-on-write rather than pickling it per worker.

\subsection{Released walking simulator}
\label{sec:setup-walking}

Benchmark~0 is the published \texttt{DNg100\_Stim} condition of the released fly walking model
\cite{pugliese2026}: a rate model over a DNg100-anchored subgraph of \texttt{manc} with
$4{,}604$ neurons, $144$ of them motor, driven by a current pulse into the descending neuron
DNg100(L) (index $31$). The behaviour and the stimulation paradigm it reproduces have a long
experimental literature in flies, from leg coordination during walking \cite{strauss1990coordination}
and the role of central pattern generators and descending signals in insect locomotion
\cite{bidaye2018six} to genetically targeted photostimulation of identified neurons
\cite{lima2005remote}. We import the pinned upstream repository (commit \texttt{faee4b0})
unchanged and drive it through a thin adapter rather than reimplementing its dynamics, so the
observable is the released model's own motor trace. The published adequacy criterion is an
oscillation score $\geq 0.5$ on the motor population after a $230$-ms transient; we keep that
threshold as the adequacy axis and use the same oscillation score as the continuous quality axis,
so the published binary criterion becomes one point on a curve rather than the whole result.

The protected interface is the stimulus neuron plus every motor neuron ($145$ neurons). The
released reduced circuit additionally keeps three interneurons (indices $277$, $617$ and
$1167$, i.e.\ IN17A001, INXXX466 and IN16B036), for $148$ neurons in total. Two properties of
the upstream code shape the protocol. The removal operator is a symmetric mask on a dense $W$, so
removal buys no runtime saving and batching is the only available parallelism; and stochastic
parameters are drawn once from the experiment seed for a fixed sample of simulations, so we hold
that sample size fixed and let replicate $k$ denote the same draw in every run.

The published $148$-neuron circuit is the primary baseline: the analysis uses its released
simulator and adequacy criterion, then expands its single point estimate into a fidelity--cost
frontier. Comparisons are against that published output and against matched internal controls; the
runtime of the original activity-weighted pruner is not inferred from its published circuit.

Search runs at $\mathrm{rtol}=10^{-4}$ so that a $400$-evaluation budget is affordable. Every
claim below is then re-scored at the published tolerance $\mathrm{rtol}=2\times10^{-6}$,
$\mathrm{atol}=5\times10^{-9}$, on $32$ held-out parameter draws taken from a different seed, all
of which are reference-valid. The two tolerances are not interchangeable and we do not mix them.

\subsection{Baselines and budget}

Random masked subsets across cost levels, greedy backward elimination, H-CABES, and adaptive
complement splitting are evaluated here. The published activity-weighted stochastic pruning
procedure is the natural comparator and is not evaluated under a matched call accounting here.
The methods actually compared receive identical mask-evaluation budgets and identical final test
replicates, so those comparisons are compute-matched by construction. On the walking model the budget is also the binding constraint on
what a black-box search can do at all: greedy eliminates one neuron per accepted step and probes
$32$ candidates per step, so a budget of $400$ evaluations buys roughly twelve accepted removals,
whereas a bundle method can trade verification depth for breadth and remove thousands. That
asymmetry is part of the result rather than a detail of the harness.

\subsection{Metrics}

Attained cost at each $(\epsilon,\rho)$; exact-frontier regret on tractable systems; frontier
hypervolume and area under the fidelity-cost curve; mask evaluations to first certified solution;
recovery of the target circuit or a certified equal-or-cheaper alternative; degeneracy statistics;
and the non-decomposability index.

\subsection{Compute}

The package used here is dependency-light and runs on CPU for the synthetic, analogue, and
$2{,}488$-neuron real-circuit experiments; the real frontier in Table~\ref{tab:real} took under ten
minutes per method on one core. The full male CNS ($176{,}422$ neurons) is a different proposition:
the sparse form is what makes it representable at all, whereas the released walking simulator ships a
dense weight matrix and needs a Linux host with an NVIDIA GPU. Consequently retained-substrate cost is
a statement about substrate, not about wall-clock
speedup, and we report measured runtime separately.

\section{Results}
\label{sec:results}

\subsection{The frontier can be estimated exactly on tractable systems}

Table~\ref{tab:exact} reports attained cost against exhaustive enumeration over three seeds per
structure, at a matched budget of $200$ mask evaluations with $n=10$.

\begin{table}[h]
\centering
\small
\begin{tabular}{@{}lrrrrr@{}}
\toprule
Structure & Exact & Greedy & H-CABES & Greedy regret & H-CABES regret \\
\midrule
Additive              & 2  & 2  & 2.42 & 0.00 & 0.42 \\
Redundant pair        & 1  & 1  & 1.08 & 0.00 & 0.08 \\
Jointly required pair & 2  & 2  & 2.42 & 0.00 & 0.42 \\
Threshold-$k$         & 3  & 3  & 3.42 & 0.00 & 0.42 \\
Inert core            & 2  & 2  & 2.42 & 0.00 & 0.42 \\
Planted blocks        & 10 & 10 & 10.00 & 0.00 & 0.00 \\
Equivalent optima     & 2  & 2  & 2.00 & 0.00 & 0.00 \\
\bottomrule
\end{tabular}
\caption{Attained cost against exhaustive enumeration, mean over three seeds. Greedy attains the
exact optimum in all $21$ case-seed combinations. H-CABES does not, missing in $11$ of $21$ and
carrying regret up to $0.42$. At this scale a full single-neuron sweep is affordable, so the greedy
special case is exact and the stochastic proposals of the hierarchy are a disadvantage.}
\label{tab:exact}
\end{table}

This table is a reversal of what we expected, and it is the more useful result. At $n=10$ a
single-neuron sweep costs ten evaluations, so greedy is effectively exact and H-CABES's sampled
bundles and capped proposals waste budget on candidates that are not the best single removal. The
hierarchy is therefore \emph{not} the right tool when exhaustive sweeps are affordable. Section
\ref{sec:results} shows where the ordering flips: once $n$ is large enough that one greedy sweep
exceeds the entire budget, the hierarchy is the only method that makes progress at all.

Separately, the interaction structure still separates cases that the optimal cost cannot: a redundant
pair and a jointly required pair both have exact cost $2$ but opposite non-decomposability. That
measurement is independent of which search method is used, and it is invisible to any method that
reports only a minimal set.

Fallback accounting makes the group-test cost explicit. Across $40$ controlled runs (five
structures, eight seeds; $n=12$, branching factor $4$), $14$ of $75$ union checks required the
one-at-a-time fallback ($18.7\%$). The rate was $0\%$ for additive, jointly-required-pair, and
inert-core systems, $72.7\%$ for redundant-pair systems, and $75.0\%$ for threshold-$k$ systems.
Thus the fallback is an interaction diagnostic rather than a routine cost of splitting: a failed
singleton terminates its branch, whereas only a failed union creates the extra serial work.

\subsection{Sparse interactions are missed by uniform sampling}

Pairwise interactions are sparse, so estimating them is itself a budgeted sampling problem. Drawing
pairs from the candidate pool recovers a positive-interaction frequency of $0.42$, against $0.14$ for
uniform random pairing at the same budget; uniform allocation is therefore wasteful when interactions
are sparse. The residual limitation is real: the planted-blocks structure still scores
$0.000$ because no scheme that ignores block structure reaches the within-block pairs.

\subsection{The method recovers a hidden circuit in a dynamical network}

Table~\ref{tab:dynamical} reports recovery in the $40$- and $60$-neuron oscillator networks, three
seeds per cell, four reference-valid replicates, and a matched budget. Every cell recovers the full
three-neuron motif with zero measured degradation, so recovery is not the discriminating quantity:
what differs is how much substrate each method keeps.

\begin{table}[h]
\centering
\small
\begin{tabular}{@{}rrrrr@{}}
\toprule
$n$ & Budget & Greedy cost & H-CABES cost & Motif recall \\
\midrule
40 & 60  & 38.0 & 19.7 & 3 of 3 \\
40 & 120 & 36.0 & 16.7 & 3 of 3 \\
40 & 240 & 33.0 & 11.7 & 3 of 3 \\
40 & 480 & 24.0 & 6.7  & 3 of 3 \\
60 & 60  & 58.0 & 43.3 & 3 of 3 \\
60 & 120 & 57.0 & 40.3 & 3 of 3 \\
60 & 240 & 55.0 & 34.3 & 3 of 3 \\
60 & 480 & 51.0 & 21.7 & 3 of 3 \\
\bottomrule
\end{tabular}
\caption{Hidden-motif recovery in a dynamical network, mean over three seeds. Mean $\dtask$ is
$0.000$ and motif recall is $1.0$ in every cell. H-CABES retains roughly half the substrate of
budgeted greedy at $n=40$ and between $12\%$ and $38\%$ less at $n=60$, and the gap widens with
budget. This is an analogue, not the biological benchmark.}
\label{tab:dynamical}
\end{table}

This is where the ordering from Table~\ref{tab:exact} reverses. At $n=10$ a full sweep was affordable
and greedy was exact; here $n=40$ and $n=60$ place a single sweep at or beyond the budget, and every
H-CABES cell beats every greedy cell at equal recovery. The result establishes two modest but
necessary facts: the pipeline runs end-to-end on a dynamical system rather than only on a set-function
oracle, and the search is not merely fitting that oracle. It does not establish superiority in
general, and the network is ours rather than published.

\subsection{The published walking circuit, certified (Benchmark~0)}
\label{sec:walking}

Both the model and the answer here come from outside this paper: a released simulator of a real
behaviour, together with a reduced circuit published alongside it. That result is a point estimate:
keep the stimulus, the $144$ motor neurons and three interneurons. What it cannot state, and we can,
is how much of the task survives each subset, and with what statistical guarantee.
Table~\ref{tab:walking} reports every subset at the published tolerance on $32$ held-out parameter
draws from a seed the search never saw.

\begin{table}[h]
\centering
\small
\begin{tabular}{@{}lrrrrrr@{}}
\toprule
Retained set & Size & $\dtask$ mean & worst draw & $\pfail$ & $\dfid$ & Adequate \\
\midrule
Interface only                       & 145 & 0.912 & 1.000 & 1.000 & 2.950 & $0/32$ \\
\textbf{Published core}              & \textbf{148} & \textbf{0.017} & \textbf{0.332} & \textbf{0.000} & 0.414 & $\mathbf{32/32}$ \\
Core $-$ 277                         & 147 & 0.912 & 1.000 & 1.000 & 2.950 & $0/32$ \\
Core $-$ 617                         & 147 & 0.912 & 1.000 & 1.000 & 2.935 & $0/32$ \\
Core $-$ 1167                        & 147 & 0.912 & 1.000 & 1.000 & 8.112 & $0/32$ \\
Core $-$ any two                     & 146 & 0.912 & 1.000 & 1.000 & 2.94 to 2.95 & $0/32$ \\
Core $+$ fourth interneuron (86)     & 149 & 0.0004 & \textbf{0.004} & 0.000 & 0.373 & $32/32$ \\
Core $+$ 73 random                   & 221 & 0.009 & --- & 0.000 & 0.417 & $32/32$ \\
Random 221 containing the three      & 221 & 0.015 to 0.024 & --- & 0.000 & 0.41 to 0.46 & $32/32$ \\
Random 221 omitting the three        & 221 & 0.912 & 1.000 & 1.000 & 2.950 & $0/32$ \\
Random three in place of the three   & 148 & 0.912 & 1.000 & 1.000 & 2.950 & $0/32$ \\
Partner swap $277\!\to\!273$          & 148 & 0.912 & 1.000 & 1.000 & 2.950 & $0/32$ \\
Partner swap $617\!\to\!1121$         & 148 & 0.912 & 1.000 & 1.000 & 2.935 & $0/32$ \\
Partner swap $1167\!\to\!1221$        & 148 & 0.912 & 1.000 & 1.000 & 8.112 & $0/32$ \\
Three partners only                  & 148 & 0.912 & 1.000 & 1.000 & 2.950 & $0/32$ \\
Core $+$ the three partners          & 151 & 0.017 & 0.332 & 0.000 & 0.414 & $32/32$ \\
\bottomrule
\end{tabular}
\caption{Held-out certification of the released walking model at the published tolerance
($\mathrm{rtol}=2\times10^{-6}$, $\mathrm{atol}=5\times10^{-9}$), $32/32$ reference-valid parameter
draws from a held-out seed. $\dtask$ is the mean one-sided degradation of the motor oscillation score,
\emph{worst draw} its maximum over the $32$ draws, and $\pfail$ the fraction of draws failing the
published adequacy threshold. Zero failures in $32$ draws bounds $\pfail$ by $0.091$ at $95\%$
confidence, below the $\rho=0.2$ used throughout. The two columns disagree in a way that matters: the
published circuit is inside $\epsilon=0.05$ on the mean but its worst draw degrades by $0.332$, so it
would fail a per-draw criterion, whereas adding interneuron $86$ brings the worst draw to $0.004$.
Repeat rows are the range over six independent random subsets; \emph{---} marks rows where only the
mean was retained.}
\label{tab:walking}
\end{table}

\begin{figure}[H]
\centering
\begin{tikzpicture}
\begin{axis}[
  width=\linewidth,
  height=6.8cm,
  ybar,
  bar width=5.5pt,
  ymin=0, ymax=1.02,
  ylabel={task degradation $\dtask$},
  ylabel style={font=\small},
  ytick={0,0.25,0.5,0.75,1.0},
  yticklabel style={font=\scriptsize},
  symbolic x coords={iface, core, m277, m617, m1167, m2set, p86, p73r, r221p, r221m, r3for3, sw273, sw1121, sw1221, coreplus},
  xtick=data,
  xticklabels={interface, core, core$-277$, core$-617$, core$-1167$, core$-$two, core$+86$, core$+73$r, 221 incl., 221 excl., 3 replaced, $277{\to}273$, $617{\to}1121$, $1167{\to}1221$, core$+$partners},
  x tick label style={rotate=52, anchor=east, font=\scriptsize},
  enlarge x limits=0.055,
  axis line style={draw=black!60},
  tick align=outside,
  legend style={font=\scriptsize, draw=black!30, at={(0.5,-0.45)}, anchor=north, legend columns=2},
  clip=false,
  every axis plot/.append style={draw=none, fill opacity=0.85}
]
\addplot[fill=red!55, draw=none] coordinates {
  (iface,0.9124) (m277,0.9124) (m617,0.9124) (m1167,0.9124) (m2set,0.9124)
  (r221m,0.9124) (r3for3,0.9124) (sw273,0.9124) (sw1121,0.9124) (sw1221,0.9124)
};
\addlegendentry{fails ($\dtask\approx0.912$, $\pfail=1.0$)}

\addplot[fill=green!55, draw=none] coordinates {
  (core,0.0168) (p86,0.0004) (p73r,0.0093) (r221p,0.0198) (coreplus,0.0168)
};
\addlegendentry{adequate ($\pfail=0$ on $32/32$ draws)}

\draw[red!70!black, densely dashed] (axis cs:iface,0.05) -- (axis cs:coreplus,0.05)
  node[pos=0.04, above, font=\scriptsize, text=red!70!black] {$\epsilon=0.05$};

\end{axis}
\end{tikzpicture}
\caption{Held-out certification of the released walking model at the published tolerance
($\mathrm{rtol}=2\times10^{-6}$, $\mathrm{atol}=5\times10^{-9}$), on $32$ reference-valid parameter
draws from a seed the search never saw. \emph{Interface} is the protected interface alone ($144$
motor neurons plus the stimulus); \emph{core} is the published reduced circuit (interface $+$ 277,
617, 1167); deletions and additions are relative to the interface and core as labelled; \emph{221
incl./excl.} are random $221$-neuron sets containing or omitting the published three; \emph{3
replaced} substitutes three random neurons; the swaps insert the same-type bilateral partner of each
published cell. Every subset missing any of the three sits at the same degradation --- removing one
is exactly as damaging as removing all three --- and every adequate subset contains all three.
Adding arbitrary neurons or the partners changes nothing measurable.}
\label{fig:walking-cert}
\end{figure}
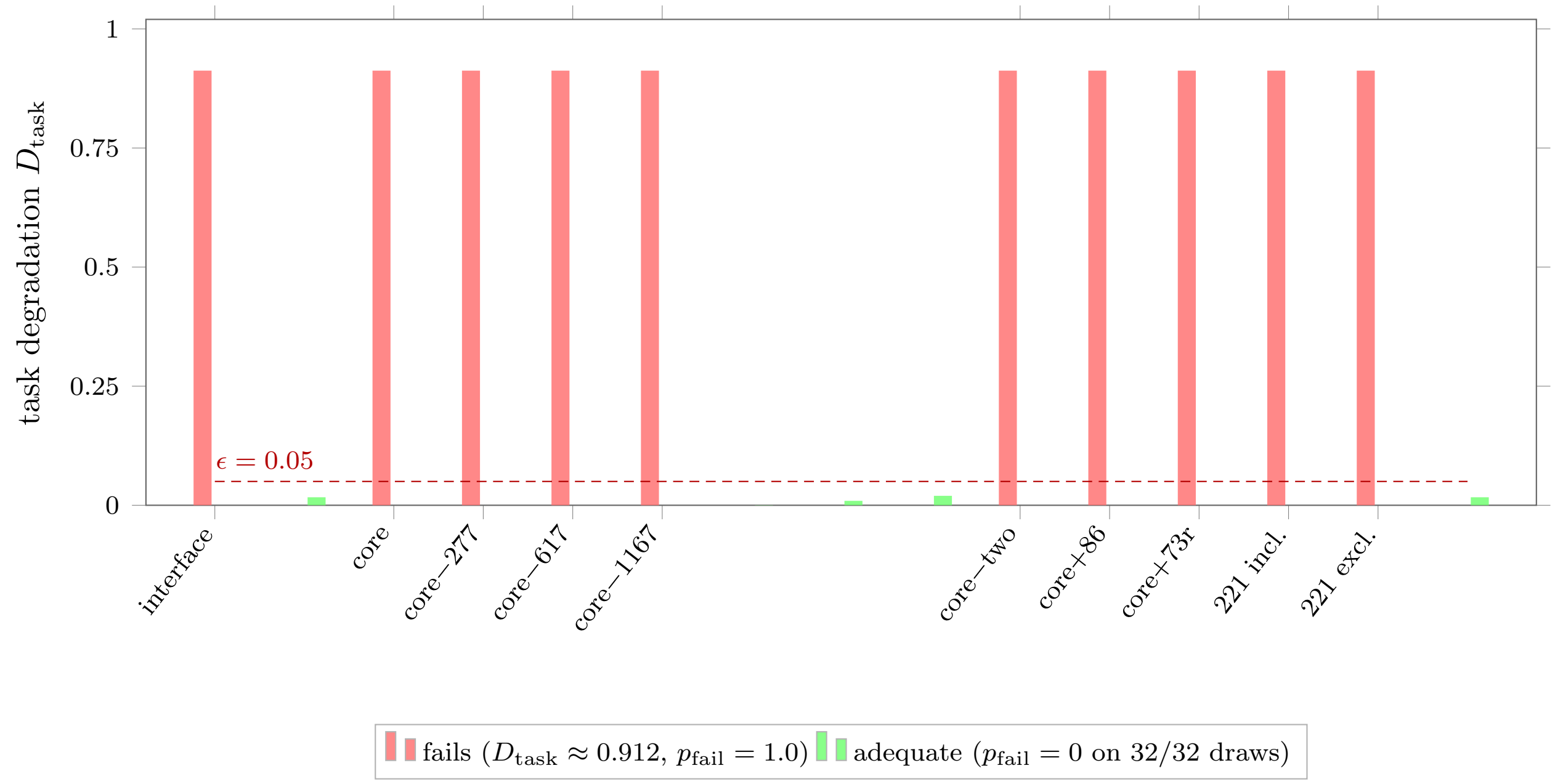


Three things follow, none of which the published point estimate states. First, the published
circuit is \emph{sufficient with margin}: $\dtask = 0.017$ against a tolerance of $\epsilon = 0.05$,
with $32/32$ draws adequate and no failure. Second, it is \emph{size-minimal within the tested
deletion class}: the interface alone degrades by $0.912$ and is adequate on none of the $32$ draws,
every one of the three single deletions and all three double deletions behave identically, so no
$146$- or $147$-neuron deletion subset of the published circuit is sufficient and $148$ is the
smallest sufficient set among those tested. The claim is restricted to deletions from the published
core: arbitrary $146$- or $147$-neuron subsets of the $4{,}604$-neuron model are not enumerated.
Third, each of the three interneurons is \emph{individually necessary within the tested deletion
class}: removing
any one is as damaging as removing all three.

The remaining rows separate two things a minimum cannot distinguish. The three cells are not
interchangeable with their own bilateral partners. Each is one member of a same-type pair
($277/273$, $617/1121$, $1167/1221$), and substituting the partner for any of them destroys the
task exactly as deleting it does, so the published side is not an arbitrary choice within a
degenerate pair. Simultaneously, \emph{adding} those same partners changes nothing measurable: the
metrics for core-plus-partners are identical to the core to three decimal places. Arbitrary
additions behave the same way: two hundred and twenty-one neurons containing the three score
$0.015$ to $0.024$ regardless of which other $73$ are chosen, while two hundred and twenty-one
neurons omitting them fail on every draw. So within the tested interventions (single deletions
from the published core, same-type partner substitutions, random supersets and random three-sets),
functional adequacy collapses onto three named interneurons, and everything else the mask retains is
inert. Global uniqueness is not claimed: the $\binom{4459}{3}$ alternative three-sets are not
enumerated.

That last point also separates the search strategies. At a matched budget of $400$ mask
evaluations, H-CABES removes $4{,}383$ of the $4{,}459$ searchable neurons ($98.3\%$) while
retaining the three necessary cells, but ends at $221$ neurons and certification shows $73$ of those
are inert. Adaptive complement splitting reaches the published $148$ in only $80$ evaluations,
then passes the strict held-out certification. The reason is structural rather than statistical. The
bundle method's cost is paid per
\emph{verification}, and greedy in particular can only eliminate what it has explicitly probed one
neuron at a time, so at this budget it spends its evaluations long before it reaches the minimum.
The published circuit was found by a domain-specific pruner. What the present numbers
support is narrower and stronger: an adaptive black-box group test recovers and certifies the
published minimum, while the original bundle hierarchy does not.

These two methods are not competitors so much as occupants of different regimes, and the contrast is
itself a finding. H-CABES is built for interaction-rich, distributed
sufficiency, where no small critical core exists and breadth of removal matters more than resolving
any particular neuron; complement splitting is built for the sparse critical-core regime, where a
few necessary neurons must be located among thousands of inert ones and each verification is
expensive. The walking benchmark is the second regime; the $40$- and $60$-neuron oscillator
networks and the whole-CNS search are the first. The framework therefore presents a toolkit rather
than a single method, and both outcomes are reported.

One row carries a different lesson. Adding the fourth interneuron (index $86$, IN19A007) reduces
$\dtask$ from $0.017$ to $0.0004$ and $\dfid$ from $0.414$ to $0.373$. It is not necessary (the
task is adequate without it), yet it is not inert either, and its contribution is invisible to any
measurement that reports only a minimal set or only a pass/fail threshold. That is the
distinction between adequacy and fidelity that the formalism insists on, appearing in a published
model rather than in a synthetic oracle.

\subsection{Recovery across seeds, and what the batching shortcut costs}
\label{sec:recovery}

\begin{figure}[H]
\centering
\begin{tikzpicture}
\begin{axis}[
  width=\linewidth, height=4.3cm,
  ymode=log, ymin=110, ymax=7000,
  ylabel={retained neurons}, ylabel style={font=\small},
  xlabel={search seed}, xlabel style={font=\small},
  symbolic x coords={0,1,2,3,4,5,6}, xtick=data,
  xticklabel style={font=\scriptsize},
  ytick={100,200,400,1000,2000,4000},
  log ticks with fixed point,
  yticklabel style={font=\scriptsize},
  axis line style={draw=black!60},
  clip=false
]
\addplot[only marks, mark=*, mark size=2.6pt, blue!70!black] coordinates {
  (0,148) (1,148) (2,319) (3,4046) (4,148) (5,148) (6,148)
};
\draw[red!70!black, densely dashed] (axis cs:0,148) -- (axis cs:6,148);
\node[font=\scriptsize, text=red!70!black, anchor=west, xshift=7pt, yshift=3pt] at (axis cs:2,148) {published size, 148};
\node[font=\scriptsize, anchor=south, yshift=1pt] at (axis cs:2,319) {319};
\node[font=\scriptsize, anchor=south, yshift=1pt] at (axis cs:3,4046) {4046};
\node[font=\scriptsize, anchor=south, yshift=1pt, text=blue!70!black] at (axis cs:0,148) {148};
\node[font=\scriptsize, anchor=south, yshift=1pt, text=blue!70!black] at (axis cs:1,148) {148};
\node[font=\scriptsize, anchor=south, yshift=1pt, text=blue!70!black] at (axis cs:4,148) {148};
\node[font=\scriptsize, anchor=south, yshift=1pt, text=blue!70!black] at (axis cs:5,148) {148};
\node[font=\scriptsize, anchor=south, yshift=1pt, text=blue!70!black] at (axis cs:6,148) {148};
\end{axis}
\end{tikzpicture}

\vspace{2mm}
\begin{tikzpicture}
\begin{axis}[
  width=\linewidth, height=3.6cm,
  ybar, bar width=10pt,
  ymin=0, ymax=430,
  ylabel={evaluations used}, ylabel style={font=\small},
  xlabel={search seed}, xlabel style={font=\small},
  symbolic x coords={0,1,2,3,4,5,6}, xtick=data,
  xticklabel style={font=\scriptsize},
  ytick={0,100,200,300,400},
  yticklabel style={font=\scriptsize},
  axis line style={draw=black!60},
  nodes near coords,
  nodes near coords style={font=\scriptsize},
  clip=false
]
\addplot[fill=blue!35, draw=blue!60!black] coordinates {
  (0,80) (1,74) (2,152) (3,89) (4,66) (5,97) (6,80)
};
\draw[black!55, densely dotted] (axis cs:0,400) -- (axis cs:6,400);
\node[font=\scriptsize, text=black!60, anchor=west, xshift=6pt] at (axis cs:2,405) {budget cap, 400};
\end{axis}
\end{tikzpicture}
\caption{Recovery of the published circuit across seven search seeds at a budget of $400$
evaluations. Five of seven seeds reach the published size of $148$ neurons and three of those recover
the published set exactly, each in $66$--$97$ evaluations --- well inside the budget, so the failures
are properties of the search path rather than of the budget. Seed~2 stalls at $319$ and seed~3 at
$4046$, which is why the headline is five of seven and not seven of seven.}
\label{fig:recovery}
\end{figure}
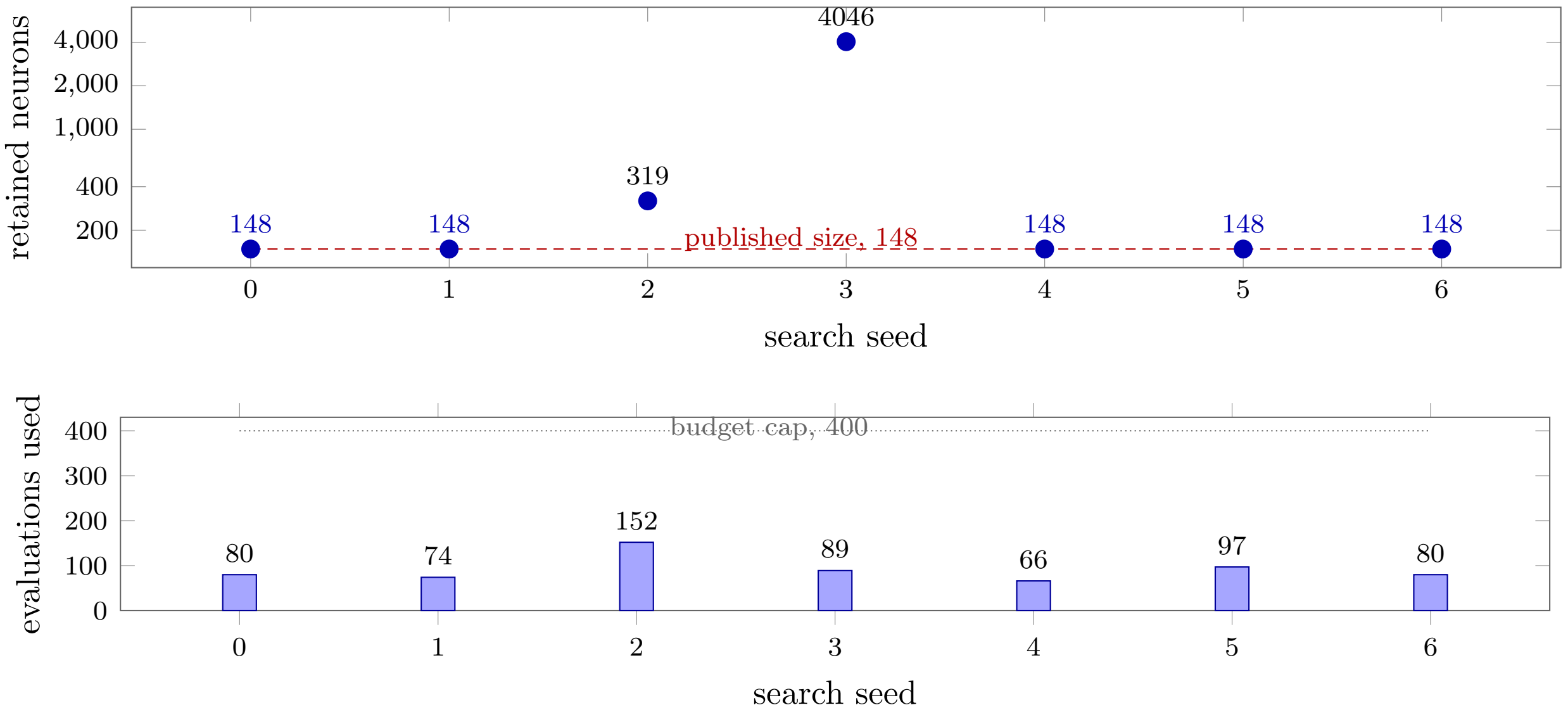

The certification above fixes the answer; it does not show that the search can find it. We therefore
ran the adaptive split from seven independent search seeds at a budget of $400$ mask evaluations,
with everything else held fixed. Figure~\ref{fig:recovery} reports the outcome. Five of the seven
seeds reach the published size of $148$ neurons, and three of those recover the published set
exactly; all five finish in $66$--$97$ evaluations, far inside the budget, so the two failures are
properties of the search path rather than of the budget. Seed~2 stops at $319$ neurons and seed~3 at
$4046$. We report five of seven rather than the best seed, because the best seed is what a single
lucky run would have shown.

\begin{table}[h]
\centering
\small
\begin{tabular}{@{}rrrrrl@{}}
\toprule
Seed & Evaluations & Retained & Removed & $\dtask$ & Recovers published set \\
\midrule
0 & 80  & \textbf{148} & 4456 & 0.0105 & \textbf{yes} \\
1 & 74  & \textbf{148} & 4456 & 0.0407 & no (drops 1167) \\
2 & 152 & 319          & 4285 & 0.0297 & no (drops 1167) \\
3 & 89  & 4046         & 558  & 0.0298 & no (stalls) \\
4 & 66  & \textbf{148} & 4456 & 0.0407 & no (drops 1167) \\
5 & 97  & \textbf{148} & 4456 & 0.0105 & \textbf{yes} \\
6 & 80  & \textbf{148} & 4456 & 0.0105 & \textbf{yes} \\
\bottomrule
\end{tabular}
\caption{Per-seed outcome of the reference search configuration at a budget of $400$ mask
evaluations, search tolerance $\mathrm{rtol}=10^{-4}$. Five seeds reach the published size, three of
them recovering the published set exactly; the two that reach $148$ without neuron $1167$ keep
$\{277, 617, 86\}$ instead. All five successful seeds finish in under a quarter of the budget, so the
failures are search-path properties, not budget exhaustion. Values are at the search tolerance used
for selection and are not the certified numbers of Table~\ref{tab:walking}.}
\label{tab:recovery}
\end{table}

The comparison between search variants is a cautionary result about the search methods themselves. The
sequential split is the reference configuration and is what Figure~\ref{fig:recovery} reports. A
sibling-batched variant, added later so that a process pool could parallelise a whole-CNS run,
evaluates the children of a failed bundle in one call and applies the accepted removals together.
It is faster per evaluation but takes a coarser step: on four seeds at the same budget it reaches
$148$ only once ($385$, $148$, $4046$, $1326$). Parallelism and search quality trade against each
other here, and the batched variant is not the one we would use on a small circuit.

\subsection{Alternative solutions and degeneracy at a fixed size}
\label{sec:degeneracy}

Because the three published cells are individually necessary, one might expect the minimum to be
unique. It is not, but the alternatives are tolerance-dependent, and certifying them is what
separates the two statements. Two seeds terminated at $148$ and $149$ neurons \emph{without}
interneuron $1167$: seed~1 kept $\{277, 617, 86\}$ and seed~3 kept a different set of four
searchable neurons. Re-scored at the published tolerance on the same $32$ held-out draws, both fail
the $\epsilon = 0.05$ gate ($\dtask = 0.078$ with $p_{\mathrm{fail}} = 0.031$, and $\dtask = 0.090$
with $p_{\mathrm{fail}} = 0$), but both clear $\epsilon = 0.10$. So at $\epsilon = 0.05$ the
published circuit is the unique certified minimum among everything we found, while at
$\epsilon = 0.10$ there are at least three distinct sufficient sets of size $148$--$149$. Degeneracy
is a property of the tolerance, not of the circuit alone, which is a distinction a single-threshold
result cannot draw.

The composition of the alternative isolates what neuron $86$ does. Removing $1167$ from the
published circuit destroys the task ($\dtask = 0.912$), but replacing it with $86$ recovers most of
the phenotype ($0.078$), and \emph{adding} $86$ to the complete published circuit improves fidelity
from $0.414$ to $0.373$ and degradation from $0.017$ to $0.0004$ (Table~\ref{tab:walking}). Neuron
$86$ is therefore neither necessary nor inert: it is a partial substitute for $1167$ and an
independent fidelity contributor. That is a role the necessary/dispensable dichotomy cannot express,
and it is present in a released model, not only in our synthetic oracles.

\subsection{Interaction structure of the published circuit}
\label{sec:interactions}

\begin{figure}[H]
\centering
\begin{tikzpicture}
\begin{axis}[
  width=0.95\linewidth,
  height=5.2cm,
  ybar,
  bar width=9pt,
  ymin=-1.0, ymax=0.30,
  ylabel={pairwise interaction $\gamma$},
  ylabel style={font=\small},
  symbolic x coords={p1, p2, p3, p4, p5, null},
  xtick=data,
  xticklabels={277--617, 277--1167, 617--1167, 86--277, 86--617, random pairs},
  x tick label style={rotate=25, anchor=east, font=\scriptsize},
  ytick={-1.0,-0.75,-0.5,-0.25,0,0.25},
  yticklabel style={font=\scriptsize},
  axis line style={draw=black!60},
  enlarge x limits=0.09,
  nodes near coords,
  nodes near coords style={font=\scriptsize, anchor=south},
  clip=false
]
\addplot[fill=blue!35, draw=blue!60!black] coordinates {
  (p1,-0.8912) (p2,-0.0575) (p3,-0.0575) (p4,-0.0999) (p5,-0.0999) (null,0.0)
};
\draw[black!55, densely dotted] (axis cs:p1,0) -- (axis cs:null,0);
\node[font=\scriptsize, anchor=north, text=blue!60!black] at (axis cs:p1,-0.95) {saturating: damage plateaus};
\node[font=\scriptsize, anchor=south, text=black!70] at (axis cs:null,0.035) {$0.000$ exactly, $24/24$ pairs};
\end{axis}
\end{tikzpicture}
\caption{Pairwise interaction structure inside the published walking circuit, from evaluations with
shared random numbers at the published tolerance. The interaction
$\gamma(i,j) = \dtask(\emptyset) + \dtask(\{i,j\}) - \dtask(\{i\}) - \dtask(\{j\})$ is the excess
damage of removing $i$ and $j$ together over the sum of the individual damages, so negative means the
two removals interfere. Neurons $277$ and $617$ show strong saturating dependence ($\gamma=-0.89$):
once either is removed the task is already destroyed, so removing both adds almost nothing. That
pattern is consistent with a serial pathway but does not establish one, and we have no independent
connectivity argument for these two cells, so we report the measurement rather than the mechanism. The
fourth interneuron $86$, which is not part of the published circuit, interacts only weakly with both.
The rightmost bar is the null: $24$ random pairs elsewhere in the network give $\gamma=0.000$ exactly,
so the measured effects are large relative to this sampled null and uniform pair sampling is extremely
inefficient in this sparse regime.}
\label{fig:interactions}
\end{figure}
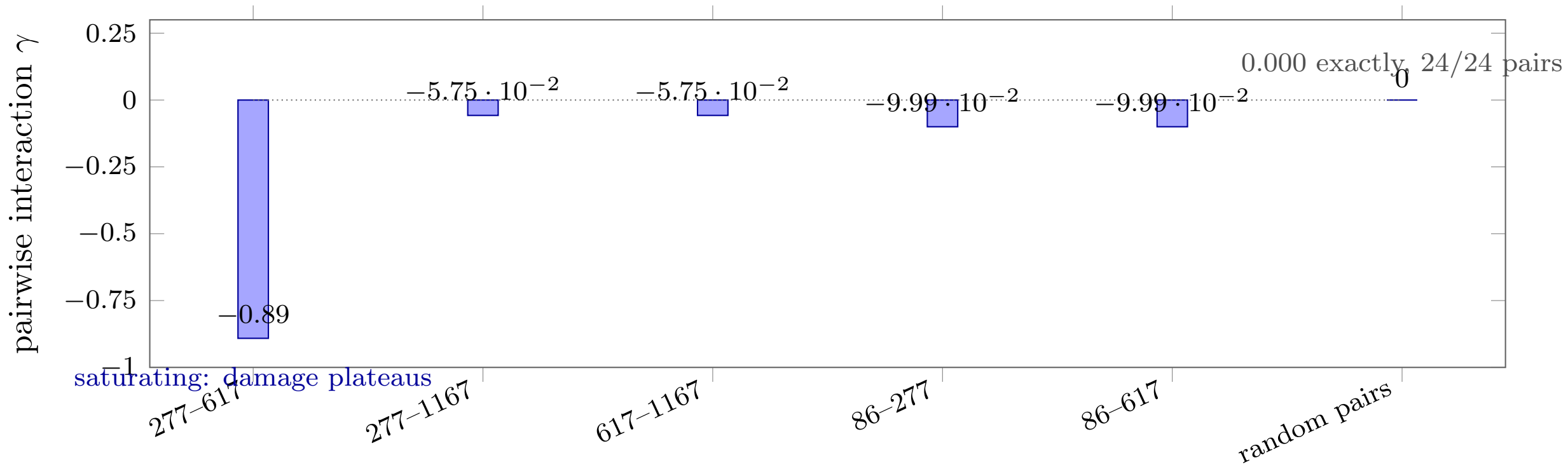

The published circuit gives the sharpest available test of non-decomposability, because exactly three
searchable neurons carry the task and their interactions can be computed rather than estimated.
Figure~\ref{fig:interactions} reports the pairwise interaction
$\gamma(i,j) = \dtask(\emptyset) + \dtask(\{i,j\}) - \dtask(\{i\}) - \dtask(\{j\})$ from real
evaluations with shared random numbers. Neurons $277$ and $617$ show strong saturating dependence
($\gamma = -0.89$): once either is removed the task is already destroyed, so removing the second adds
almost no further damage. A negative interaction of this form is consistent with a serial pathway but
does not by itself establish one, and we do not have an independent connectivity argument for these
two cells, so we describe the measurement rather than the mechanism. The remaining pairs are weakly
negative ($\gamma \approx -0.06$ to $-0.10$), and the third-order term is small and positive
($+0.058$), so the circuit is close to pairwise-decomposable with one dominant saturating pair. The
null is exact: $24$ random pairs elsewhere in the network give $\gamma = 0.000$, because removing
inert neurons changes nothing. The effects we report are therefore large relative to this sampled
null, and uniform pair sampling is extremely inefficient in this sparse regime: with enough draws
it would eventually find them, but the sampling cost is prohibitive.

\subsection{The frontier depends on the task}
\label{sec:task-conditioning}

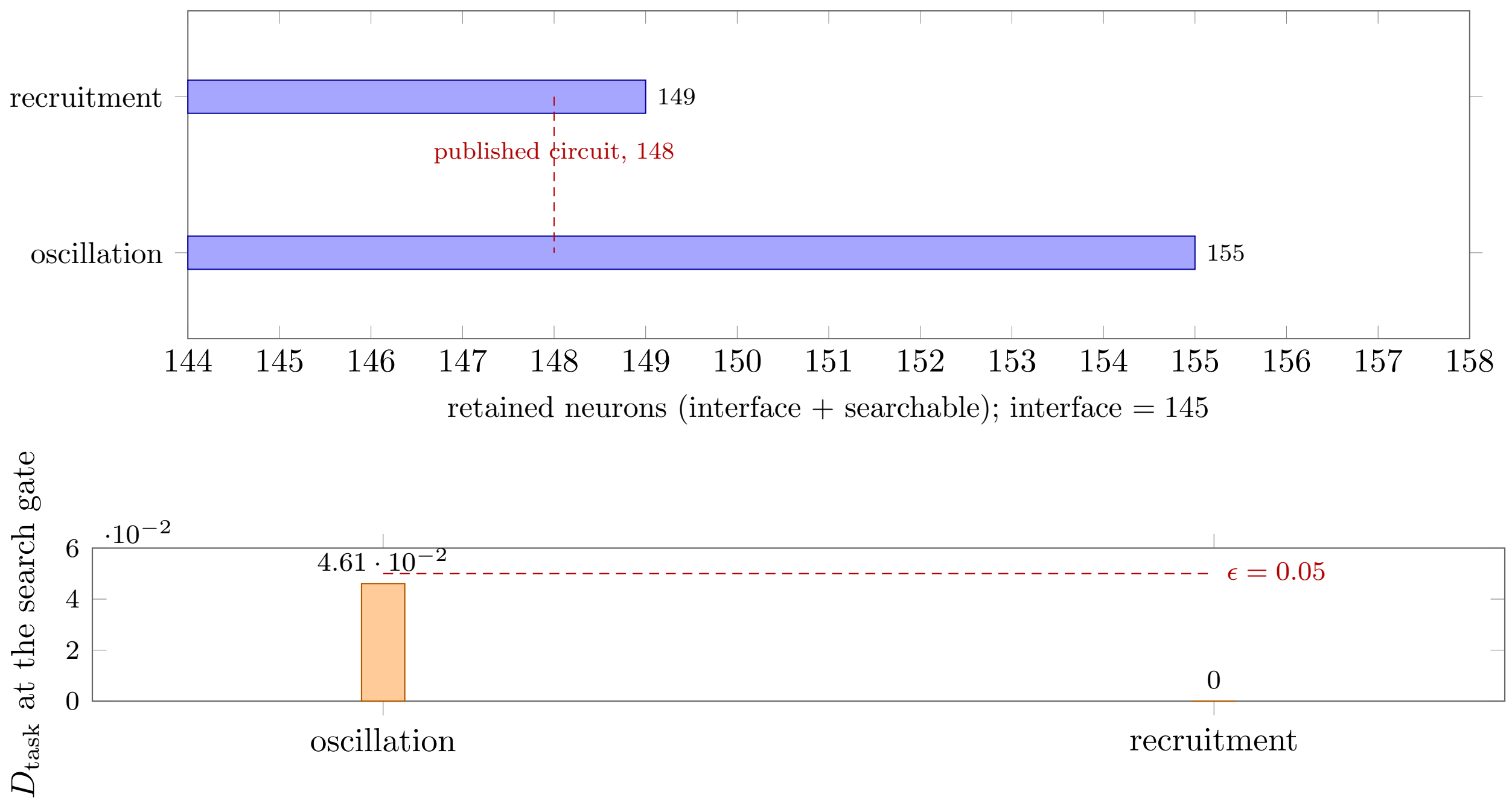
\begin{figure}[H]
\centering
\resizebox{\linewidth}{!}{%
\begin{tikzpicture}
\begin{axis}[
  width=\linewidth,
  height=5.4cm,
  xbar,
  bar width=11pt,
  xmin=144, xmax=158,
  xlabel={retained neurons (interface $+$ searchable); interface $=145$},
  xlabel style={font=\small},
  ytick=data,
  yticklabel style={font=\small},
  symbolic y coords={recruitment, oscillation},
  y dir=reverse,
  nodes near coords,
  nodes near coords style={font=\scriptsize, anchor=west},
  axis line style={draw=black!60},
  enlarge y limits=0.55,
  clip=false
]
\addplot[fill=blue!35, draw=blue!60!black] coordinates {(155,oscillation) (149,recruitment)};
\draw[red!70!black, densely dashed] (axis cs:148,recruitment) -- (axis cs:148,oscillation)
  node[pos=0.5, above, font=\scriptsize, text=red!70!black, rotate=0] {published circuit, 148};
\end{axis}
\end{tikzpicture}%
}
\vspace{-2mm}
\resizebox{\linewidth}{!}{%
\begin{tikzpicture}
\begin{axis}[
  width=\linewidth,
  height=3.2cm,
  ybar, bar width=13pt,
  ymin=0, ymax=0.06,
  ylabel={$\dtask$ at the search gate},
  ylabel style={font=\small},
  symbolic x coords={oscillation, recruitment},
  xtick=data,
  xticklabel style={font=\small},
  ytick={0,0.02,0.04,0.06},
  yticklabel style={font=\scriptsize},
  axis line style={draw=black!60},
  nodes near coords,
  nodes near coords style={font=\scriptsize},
  enlarge x limits=0.35,
  clip=false
]
\addplot[fill=orange!40, draw=orange!70!black] coordinates {(oscillation,0.0461) (recruitment,0.0000)};
\draw[red!70!black, densely dashed] (axis cs:oscillation,0.05) -- (axis cs:recruitment,0.05);
\node[font=\scriptsize, text=red!70!black, anchor=west] at (axis cs:recruitment,0.051) {$\epsilon=0.05$};
\end{axis}
\end{tikzpicture}%
}
\caption{Task conditioning on the released walking model. The simulator, protected interface, cohort,
search method and evaluation budget are identical; only the phenotype changes. Top: best-found solution
size, against the published circuit's $148$. Bottom: the degradation each solution achieves at the
search gate. Because the $145$-neuron interface is shared by construction, the comparison is made on
the searchable parts, which overlap with Jaccard $0.273$. Both runs gate at $\epsilon=0.05$, not the
$\epsilon=0.35$ of Section~\ref{sec:recovery}, so these sizes are comparable across tasks but not to
the $148$ recovered there. Sizes are best-found, not proven minimal.}
\label{fig:tasks}
\end{figure}

The word \emph{task-conditioned} is the central claim of the formalism, and until now the released
walking model contributed exactly one task: the published oscillation criterion. We therefore held
the simulator, the protected interface, the cohort construction, the search method and the budget
fixed and varied only the phenotype. Figure~\ref{fig:tasks} reports the outcome for the two tasks
that completed. Our best-found solution for the published oscillation criterion, which requires the
mean oscillation score over active motor neurons to reach $0.5$, retains $155$ neurons; our best-found
solution for a breadth task, which requires half of the intact fraction of motor neurons to oscillate
individually, retains $149$. The comparison that matters is between the searchable parts, since the
$145$-neuron interface is identical by construction. Those parts share only $3$ of their $11$ distinct
neurons (Jaccard $0.273$), with seven neurons unique to the oscillation solution and one, neuron
$696$, unique to the breadth solution. The two sets are therefore not nested in either direction, and
the same circuit, evaluated on the same motor traces, needs largely different neurons for different
tasks.

The difference is robust to the choice of search variant. The batched variant, which as
Section~\ref{sec:recovery} shows is the weaker search, independently returned exactly the same
recruitment set (Jaccard $1.00$) and the same oscillation size. Since a method whose error is large
on this problem still reproduces the ordering, the ordering is a property of the task rather than of
the search.

Two caveats belong with Table~\ref{tab:tasks} and neither is cosmetic. First, these runs gate the search at
$\epsilon = 0.05$ while the recovery experiments gate at $\epsilon = 0.35$, so the values are
comparable \emph{across tasks} but not to the $148$ of Section~\ref{sec:recovery}; the smaller number
reflects a looser gate, not a better circuit. Second, the two tasks are not equally demanding:
the breadth task is satisfied by roughly one oscillating motor neuron, so its frontier being lower
is what one expects. The result establishes that the conditioning is real; it does not establish that
it is large.

\begin{table}[h]
\centering
\small
\begin{tabular}{@{}>{\raggedright\arraybackslash}p{3.2cm}rrr>{\raggedright\arraybackslash}p{4.8cm}@{}}
\toprule
Task & Retained & Searchable part & $\dtask$ & Searchable neurons retained \\
\midrule
Oscillation (published) & 155 & 10 & 0.0461
  & 86, 277, 280, 485, 617, 630, 749, 2309, 2880, 3005 \\
Recruitment breadth & 149 & 4 & 0.0000
  & 277, 280, 617, 696 \\
\midrule
\multicolumn{5}{@{}l@{}}{\emph{Breadth solution reproduced under the independent search variant}}\\
Breadth (batched) & 149 & 4 & 0.0000 & identical (Jaccard $1.000$) \\
\bottomrule
\end{tabular}
\caption{Task conditioning on the released walking model with the simulator, protected interface,
cohort, search method and budget held fixed. The $145$-neuron interface is identical in every row, so
overlap is computed on the searchable parts alone. Those parts share only three neurons
(Jaccard $0.273$): seven are unique to the oscillation solution and neuron $696$ is unique to the
breadth solution, so neither set contains the other. Both drop interneuron $1167$ under this gate. The
breadth solution is reproduced exactly by the sibling-batched variant, which is a materially weaker
search on this problem (Table~\ref{tab:recovery}), so the difference between tasks is not an artefact
of the search. Both runs gate at $\epsilon=0.05$ rather than the $\epsilon=0.35$ of
Section~\ref{sec:recovery}, so these sizes are comparable to each other but not to the $148$ of
Table~\ref{tab:recovery}. Sizes are best-found, not proven minimal.}
\label{tab:tasks}
\end{table}

\subsection{A real ventral-nerve-cord circuit}

We instantiate the framework on a circuit retrieved from the male adult nerve cord
(\texttt{manc:v1.2.1}) through neuPrint: a DNg100-anchored subgraph of $2{,}488$ neurons and
$299{,}069$ synapses, with $72$ protected interface neurons (the stimulated seeds and $70$ motor
readouts). The dynamics are the imposed rectified-tanh rate model of Section~\ref{sec:setup} with
row-normalised weights and Gaussian gain noise; the target is the intact motor response and both
adequacy and fidelity are measured against it. These quantities are
properties of this model, not of the animal.

\paragraph{The fixed-scale configuration measured the budget.}
A bundle size fixed at three, with the search capped at $150$ evaluations, reports
$\Cstar = 2487$ --- exactly one neuron removed --- at every $\epsilon\in[0.05,0.7]$. The
cause is arithmetic, not biological: a bundle of three removes $0.1\%$ of the substrate per accepted
move, and scoring every one of the $2{,}416$ searchable neurons for a \emph{single} greedy step
already exceeds the whole budget, so a fixed bundle size cannot complete its first greedy step. The
apparent null is therefore a property of the budget rather than of the substrate: with the
scale-aware proposals of Section~\ref{sec:method} the same budget traces a graded frontier
(Table~\ref{tab:real}), and the same trap applies to any pruner whose step size is fixed while the
search space grows.

\begin{table}[h]
\centering
\small
\begin{tabular}{@{}rrrrrr@{}}
\toprule
$\epsilon$ & Cost & Removed & $\dtask$ (search) & $\dtask$ upper (held-out) & Certified \\
\midrule
0.05 & 2487 & 1   & $7.7\times10^{-5}$ & $4.0\times10^{-5}$ & \checkmark \\
0.10 & 2464 & 24  & 0.056              & 0.036              & \checkmark \\
0.20 & 2393 & 95  & 0.195              & 0.169              & \checkmark \\
0.40 & 2346 & 142 & 0.333              & 0.311              & \checkmark \\
0.50 & 2212 & 276 & 0.490              & 0.471              & \checkmark \\
\bottomrule
\end{tabular}
\caption{MANC DNg100 subgraph, budget $400$ mask evaluations. Cost is retained neurons out of
$2{,}488$. Upper bounds are one-sided bootstrap bounds on $40$ held-out replicates disjoint from
search; every row also passes the failure bound. The frontier is graded, and it replaces a flat
``one neuron removable'' line that the fixed-scale configuration produced at all five tolerances.}
\label{tab:real}
\end{table}

\paragraph{Matched budget separates the search from backward elimination.}
At an identical budget of $400$ mask evaluations and identical final certification, budgeted greedy
backward elimination ($p=16$ probes per step) removed $25$ neurons at \emph{every} tolerance
($\dtask = 2.6\times10^{-4}$), while H-CABES removed $276$ at $\epsilon=0.5$. The gap is a factor of
eleven in removed substrate at equal cost. Backward elimination is not a weak method in general
(it is a special case of ours), but at this scale its $O(n)$-per-move cost is fatal, which is why
move granularity, not targeting, is what the budget buys here.

\paragraph{The search does not beat random removal, and we report it.}
We compared each frontier point against uniformly random removals of the same size
(Table~\ref{tab:random}). The two are indistinguishable: at $24$, $95$, $142$ and $276$ neurons
removed, random removal degrades the response by $0.052$, $0.207$, $0.309$ and $0.502$, against
$0.056$, $0.195$, $0.333$ and $0.490$ for the search. On this circuit and this model, degradation is
close to a function of \emph{how many} neurons are removed rather than \emph{which}, so no privileged
compact sufficient subcircuit exists to be found. The one axis on which the search does separate is
the failure constraint: random removal of $435$ or more neurons fails adequacy in every replicate,
whereas the certified search mask at $276$ removals has an upper failure bound of $0.387$ and an
observed failure rate near zero. The advantage of the method here is robustness to parameter draws,
not the discovery of a smaller circuit.

\begin{table}[h]
\centering
\small
\begin{tabular}{@{}rrrr@{}}
\toprule
Neurons removed & $\dtask$ (search) & $\dtask$ (random, mean of 6) & $\pfail$ (random) \\
\midrule
24  & 0.056 & 0.052 & 0.00 \\
95  & 0.195 & 0.207 & n/a \\
142 & 0.333 & 0.309 & n/a \\
218 & n/a   & 0.474 & 0.21 \\
276 & 0.490 & 0.502 & n/a \\
435 & n/a   & 0.579 & 1.00 \\
\bottomrule
\end{tabular}
\caption{Search versus random removal at matched removal count. n/a marks sizes not sampled on that
curve. Within the noise of six draws per level the search and random removal are indistinguishable
on degradation; they separate on the failure rate.}
\label{tab:random}
\end{table}

\paragraph{The circuit is not more non-decomposable than its rewired control.}
We sampled $60$ neuron pairs and computed the signed pair interaction
$\gamma_{ij}=\dtask(\{i,j\})-\dtask(\{i\})-\dtask(\{j\})$. The non-decomposability index on the real
circuit is $2.2\times10^{-4}$; on an out-degree-preserving rewiring of the same graph it is
$3.7\times10^{-4}$, larger. The absolute values are negligible against the $\dtask$ scale of
$0.1$ to $0.5$. In this model the wiring is therefore \emph{not} more irreducibly
higher-order than a degree-matched random graph, the opposite of what a modular-prior expectation
would suggest. Because this model is all-excitatory, that is a statement about the model rather than
about the fly nerve cord, and the CNS result below supersedes it; it is reported as a negative result
rather than as an absence of data.

\subsection{The complete male CNS, with real inhibition}

The ventral-nerve-cord result above is limited by a defect that the full CNS does not share: that
circuit carried
no transmitter annotation, so the model was entirely excitatory and its only stable regime was
saturation. The complete male CNS (\texttt{male-cns:v1.0}) does carry transmitter predictions, and
the difference is decisive.

\paragraph{Scale and signs.}
We pull the whole dataset through neuPrint: $176{,}422$ neurons and $25{,}862{,}574$ connections
after per-neuropil weights are summed, mean out-degree $146.6$. \texttt{predictedNt} gives
$127{,}866$ excitatory and $48{,}556$ inhibitory (GABA or glutamate) neurons, so $27.5\%$ of the
network is inhibitory, against $0\%$ in the earlier model. The stimulated input population is the
whole sensory class ($17{,}937$ neurons) and the readout is every motor neuron ($815$); these
$18{,}752$ neurons are the protected interface and all other $157{,}670$ are searchable. Weights are
not row-normalised, so degree and weight structure survive; a global scale of $1/\bar d$ and a tonic
bias of $0.55$ place the network in a graded, non-saturated regime, chosen before the experiment by
the criterion that removal be neither silent nor saturating.

\paragraph{A sparse task-supporting subgraph suffices in the model.}
Table~\ref{tab:cns} reports the certified frontier at $1200$ mask evaluations per method. Removing
$62{,}844$ neurons, $40\%$ of the searchable network, degrades the motor response by $0.0069$,
which is essentially nothing, and is certified with an upper bound of $0.0114$ on $30$ held-out
replicates. Removing $87{,}987$ costs $0.082$. At the largest tested removal, the retained set is the
$18{,}752$-neuron protected interface plus $93$ neurons and reaches $0.462$ at $\epsilon=0.5$.

This is an AI-model statement about effective computation in a weighted graph, not a claim that
the corresponding fraction of biological synapses is dispensable. In the imposed rate model, the task can be
carried by a small retained subgraph; strong effective weights and short relay paths are a plausible
mechanistic interpretation of that concentration of useful drive, but path attribution was not
measured here and no biological mechanism is inferred.

\begin{table}[h]
\centering
\small
\begin{tabular}{@{}rrrrr@{}}
\toprule
$\epsilon$ & Cost & Removed & $\dtask$ (search) & Held-out certified \\
\midrule
0.05 & 113{,}578 & 62{,}844  & 0.0069 & \checkmark\ ($\leq 0.0114$) \\
0.10 & 88{,}435  & 87{,}987  & 0.0815 & no ($0.1020$) \\
0.20 & 88{,}435  & 87{,}987  & 0.0815 & \checkmark\ ($\leq 0.1020$) \\
0.30 & 88{,}435  & 87{,}987  & 0.0815 & \checkmark\ ($\leq 0.1020$) \\
0.40 & 22{,}177  & 154{,}245 & 0.3838 & no ($0.4442$) \\
0.50 & 18{,}845  & 157{,}577 & 0.4624 & \checkmark\ ($\leq 0.4661$) \\
0.60 & 18{,}845  & 157{,}577 & 0.4624 & \checkmark\ ($\leq 0.4661$) \\
\bottomrule
\end{tabular}
\caption{Complete male CNS, budget $1200$ mask evaluations, $30$ held-out replicates. Cost is
retained neurons of $176{,}422$. Certification is a one-sided bootstrap bound on held-out damage;
two rows fail narrowly, which is the bound doing its job rather than a defect.}
\label{tab:cns}
\end{table}

\paragraph{The interface alone fails, so the frontier is not vacuous.}
A frontier that removes almost everything is worthless if the protected interface already solves
the task, so we measure that baseline directly: retain $18{,}752$ interface neurons, delete all
$157{,}670$ searchable ones. The interface alone reaches $\dtask = 0.746$ (a response-match score
of $0.254$) and fails adequacy in every replicate ($\pfail = 1.0$). It is not sufficient. The
$\epsilon=0.5$ solution retains $18{,}845$ neurons, which is the interface plus \emph{ninety-three}
neurons, and reaches a certified $0.462$. Those $93$ neurons, selected from $157{,}670$ candidates,
convert a mask that fails every replicate into one that passes. Those ninety-three neurons matter
more than the removal fraction suggests.

For scale: random retention of the smallest level we sampled, $788$ searchable neurons, eight
times as many, reaches only $0.621$ at best and $0.746$ on average. The search reaches $0.462$
with $93$.

\paragraph{Targeted search materially outperforms uniform removal.}
At the identical budget, budgeted greedy backward elimination removed $75$ neurons at \emph{every}
tolerance, against $157{,}577$ for H-CABES at $\epsilon=0.5$. The gap is a factor of $2{,}100$, and
it is not a tuning artefact: greedy's $O(n)$-per-move cost cannot complete enough single-neuron
steps inside the budget at $n=176{,}422$.

\paragraph{Uniform removal does not substitute for targeted search.}
The control that decides this is random removal of the same size (Table~\ref{tab:cns-random}, $240$ masks).
Removing $62{,}844$ neurons at random degrades the response by $0.727$ on average, and by no better
than $0.304$ in the best of twelve draws, where the search achieves $0.0069$, a factor of about
$105$ at this operating point. At $157{,}577$ removed the random mean is $0.746$ and the best draw
$0.621$, against the search's certified $0.462$, a margin of about $1.6\times$: the advantage
narrows as the search approaches its floor. Random
removal of $74{,}313$ or more fails adequacy in \emph{every} replicate; the search's
$157{,}577$-removal mask carries an upper failure bound of $0.394$. Sufficiency here depends on which
neurons are retained, not on how many: the search finds structure that uniform deletion destroys.

\begin{table}[h]
\centering
\small
\begin{tabular}{@{}rrrr@{}}
\toprule
Removed & Search $\dtask$ & Random mean (12 draws) & Random best of 12 \\
\midrule
62{,}844  & 0.0069 & 0.727 & 0.304 \\
87{,}987  & 0.0815 & 0.767 & 0.549 \\
154{,}245 & 0.3838 & 0.743 & 0.600 \\
157{,}577 & 0.4624 & 0.746 & 0.621 \\
\bottomrule
\end{tabular}
\caption{Search versus random removal at matched removal count on the full CNS. The search advantage
is largest at the shallow operating point ($0.0069$ vs $0.727$, about two orders of magnitude) and
narrows to about $1.6\times$ at $157{,}577$ removals, where the search approaches the interface
floor.}
\label{tab:cns-random}
\end{table}

\paragraph{The real wiring is more interaction-structured than its control, weakly.}
For $80$ sampled pairs, the non-decomposability index is $1.27\times10^{-8}$ on the real graph and
$1.52\times10^{-14}$ on an out-degree-preserving rewiring, a ratio of about $10^{6}$ that we report
as secondary. The effect size is what matters: both indices sit more than six orders of magnitude
below the $0.1$ to $0.5$ damage scale, so pair interactions are not what carries the sufficiency, and
at this sample size the ordering is weak evidence rather than a substantial result. It nonetheless
reverses the direction of the all-excitatory result: on that model the rewired graph looked
\emph{more} non-decomposable, and once real inhibition is present the real graph does.

\subsection{Adaptive splitting at whole-CNS scale: a negative result}
\label{sec:cns-split}

Group testing should in principle suit the CNS, because its cost scales with the size of the
necessary set rather than with the size of the network, and the H-CABES result suggests the
necessary set beyond the interface is small. We implemented a wave-batched variant of the adaptive
split that drains the whole pending frontier in one batched call per wave, which is the only way the
process pool can help: in its original form every queued part costs one sequential simulation
regardless of how many workers exist, and an earlier attempt spent three hours with sixteen idle
processes before we diagnosed it.

\begin{figure}[H]
\centering
\begin{tikzpicture}
\begin{axis}[
  width=\linewidth,
  height=5.6cm,
  xlabel={adequacy tolerance $\epsilon$},
  ylabel={retained neurons (thousands)},
  xlabel style={font=\small},
  ylabel style={font=\small},
  xmin=0.02, xmax=0.62,
  ymin=0, ymax=185,
  xtick={0.05,0.1,0.2,0.3,0.4,0.5,0.6},
  xticklabel style={font=\scriptsize},
  yticklabel style={font=\scriptsize},
  legend style={font=\scriptsize, draw=black!30, at={(0.98,0.97)}, anchor=north east},
  axis line style={draw=black!60},
  grid=major,
  grid style={draw=black!12},
  clip=false
]
\addplot[mark=*, thick, blue!70!black] coordinates {
  (0.05,113.578) (0.10,88.435) (0.40,22.177) (0.50,18.845) (0.60,18.845)
};
\addlegendentry{H-CABES, budget $1200$}
\addplot[mark=square*, thick, red!70!black, densely dashed] coordinates {
  (0.05,18.752) (0.60,18.752)
};
\addlegendentry{interface alone ($=18{,}752$) fails}
\addplot[mark=triangle*, thick, orange!80!black] coordinates {
  (0.05,176.422) (0.10,176.422) (0.20,114.831) (0.30,68.021) (0.40,58.028) (0.60,58.028)
};
\addlegendentry{wave-batched split, budget $250$}
\end{axis}
\end{tikzpicture}
\caption{The complete male CNS ($176{,}422$ neurons, $25{,}862{,}574$ connections). H-CABES reaches
$18{,}845$ neurons at $\epsilon=0.5$, only $93$ above the protected interface, and the interface
alone fails every draw, so those $93$ neurons are doing the work. The wave-batched adaptive split is
plotted as the negative result it is, and the comparison is not budget-matched: the split was capped
at $250$ evaluations against H-CABES's $1200$, so the two curves are not directly comparable and we
do not claim the split is the weaker method in general. Its only strict certificate is at
$\epsilon=0.05$ on the \emph{unmodified} network, which removes nothing and is therefore degenerate,
not a finding.}
\label{fig:cns-frontier}
\end{figure}
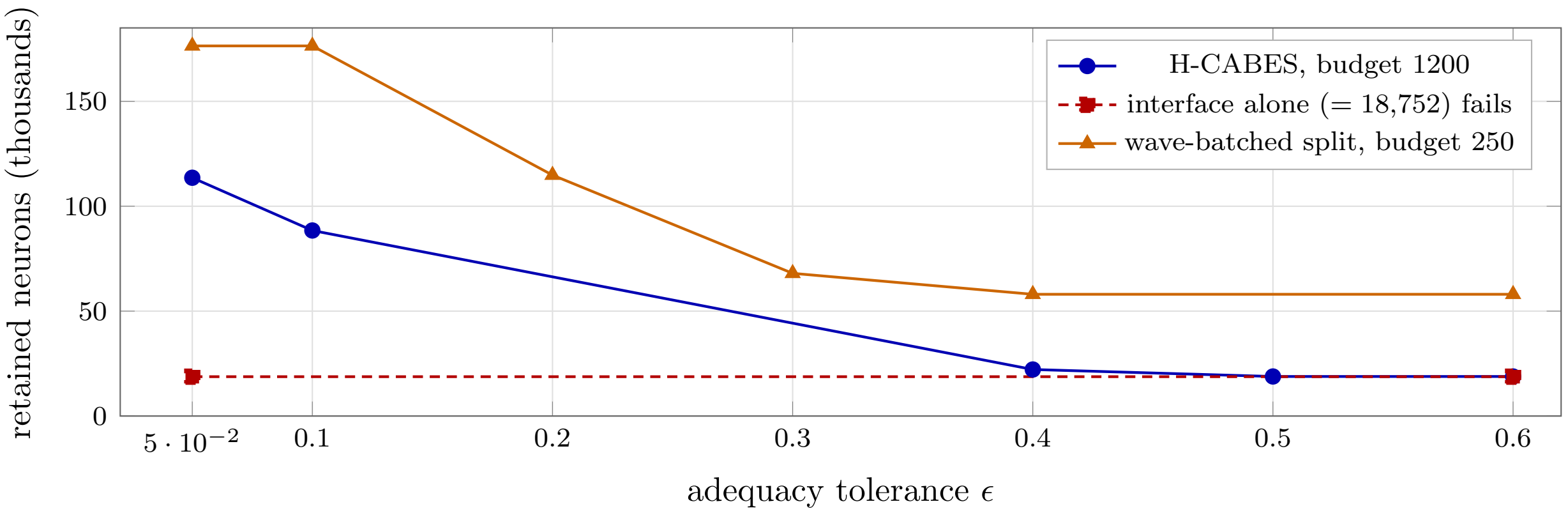

Figure~\ref{fig:cns-frontier} reports what happened, and it is a negative result for the method. The
split is dominated by H-CABES at every tolerance, and the comparison is not budget-matched (the
split was capped at $250$ evaluations against H-CABES's $1200$), so we do not conclude that the
split is the weaker method in general, only that this implementation of it is weaker at this budget
on this circuit. Its one strict certificate is on the \emph{unmodified} network at
$\epsilon = 0.05$: with nothing removed, the intact circuit trivially certifies, which is a
degenerate result and not a finding. It is included because a certificate that passes by doing nothing
is exactly the failure mode the interface-only baseline exists to catch, and a table that omitted it
would look like a success.

\subsection{What actually decides the result}

We can isolate the cause, because we re-ran the \emph{same} $2{,}488$-neuron VNC circuit under the
sign-aware model: that dataset's own \texttt{predictedNt} ($1{,}463$ inhibitory, $1{,}025$
excitatory), unnormalised weights at $10\times$ coupling, where removing $5\%$ of inhibitory neurons
damages the response $3.5\times$ more than removing $5\%$ of excitatory ones. Structure is clearly
detectable in that model. The search nevertheless removes all $2{,}416$ searchable neurons at
$\dtask = 0.051$, and it looks indistinguishable from random removal. The protected interface is
two seed neurons plus seventy motor neurons, and that interface \emph{alone} already reproduces
$0.949$ of the intact response. There is nothing left to find.

The CNS is the opposite case, and what decides the outcome is headroom rather than inhibition. Its
$18{,}752$-neuron interface reaches only $0.254$ of intact and fails every replicate, so the
remaining $157{,}670$ neurons genuinely have to be searched, and the search beats random retention by
a wide margin. The earlier all-excitatory run sits between the two: its interface reached $0.369$.

The rule this yields is that a sufficiency frontier is uninterpretable without the interface-only
baseline reported next to it. A large removable fraction is impressive only when the interface is
insufficient on its own; when the interface already suffices, ``almost everything is removable'' is
a statement about how the interface was defined rather than about the circuit. Both real-circuit
results are therefore reported with the baseline, and the CNS numbers are evidence about this model of
the male CNS, not about the animal.

The walking benchmark is the third instance of the rule and the only one where the interface has
\emph{negative} headroom. Its $145$-neuron interface degrades by $0.912$ and is adequate on none of
$32$ held-out draws, so the interface is not merely insufficient but catastrophic, and the task lives
entirely in the three neurons that the published circuit adds. That is why the removable fraction is
so high ($4{,}456$ of $4{,}459$ searchable neurons), and it is also why the number means
something: with the interface baseline reported, the reduction is attributable to three named cells
rather than to the definition of the interface. The task axis then qualifies the rule further. If
headroom is what decides the outcome, and the task determines what counts as headroom, then the same
circuit has different headroom for different tasks: the oscillation task needs ten searchable neurons
where the breadth task needs four, on identical motor traces. Headroom is not a property of the
interface alone; it is a property of the interface-task pair.

\section{Discussion}

The framing here follows directly from the state of the literature. Once one paper reports a minimal
sufficient circuit and another reports the cells and synapses that sustain an ongoing computation,
discovering one more compact
circuit is no longer a contribution. What remains open is the structure of the question: how
substrate requirements vary with tolerated error, whether the computation decomposes into independent
units, how many distinct sufficient circuits exist, and which neurons occupy which functional role.
These are properties of a set-valued object, and they require a different measurement apparatus than
a single minimum.

Two design decisions follow from anticipating reviewer objections rather than from convenience. The
failure constraint exists because mean distortion can hide rare catastrophic runs. The protected
interface exists because otherwise sufficiency can be achieved by deleting the task. Both are cheap,
and their absence would be fatal.

\section{Limitations}
\label{sec:limitations}

Sufficiency here is relative to a specified simulator, intervention operator, phenotype, evaluation
horizon, input distribution, cost, and search budget. It is not biological necessity, and a
model-sufficient circuit is not evidence that an animal needs it. Every number in this paper is a
property of a simulator under sampled parameters; the framework is what we would defend, and the
biology remains to be established.

\paragraph{The ventral-nerve-cord model has no inhibition.}
The retrieved VNC circuit carries no transmitter annotation, so that model is entirely excitatory, and
an all-excitatory rate network has saturation as its only stable regime. That makes the response
smooth and homogeneous and is the most likely explanation for the flat degradation curve and for the
search failing to beat random removal, so R3 is a property of this model rather than of the fly nerve
cord. The complete male CNS does carry transmitter predictions and the model uses them ($27.5\%$
inhibitory); the two results disagree because of the sign model alone, and the CNS result supersedes
the VNC one. The protected interface compounds the VNC defect: removing all $2{,}416$ searchable
neurons still leaves a response-match score of $0.369$, so the directly stimulated and readout
populations alone account for roughly a third of the intact response and bound the maximum attainable
degradation near $0.63$.

\paragraph{Sizes are best-found, not proven minimal.}
The frontier we compute is \bestfound{}; no global optimality is claimed. Our own search is unreliable
on the walking circuit: adaptive complement splitting reaches the published size in five of seven
seeds and the sibling-batched variant in one of four, and the two failing seeds stall well above the
minimum rather than converging, so a single reported run would overstate the method. Sufficient
circuits are generally non-unique, so recovering one circuit does not establish uniqueness, and
Jaccard overlap against a single reference circuit is not a valid primary success criterion. On the
CNS the split comparison is not budget-matched ($250$ evaluations against H-CABES's $1200$), so the
dominance in Figure~\ref{fig:cns-frontier} is not attributable to the algorithm alone.

\paragraph{Task conditioning rests on two phenotypes.}
The two tasks differ in difficulty, so the ordering between their frontiers is interpretable while
their magnitudes are not comparable, and the two runs gate at a different tolerance from the recovery
experiments ($\epsilon=0.05$ versus $\epsilon=0.35$). Parameter realisations materially affect
dynamics, so a conclusion about a computation requires agreement across a parameter ensemble or a
worst-case bound; the only robustness axis reported here is search-seed variation, which is a property
of the search rather than of the substrate. The certified failure bound is also one-sided: $0$
failures in $32$ draws bounds $\pfail$ by $0.091$ at $95\%$ confidence, which excludes $\rho=0.2$ but
would not exclude a failure rate of a few percent. Larger held-out cohorts would tighten it.

\paragraph{The comparisons are internal.}
The activity-weighted stochastic pruning procedure that produced the walking circuit is not evaluated
under a matched call accounting, so every search comparison is between the methods introduced here and
random controls rather than against the procedure that solved the benchmark. Since those methods
disagree with each other by wide margins depending on regime, such a comparison could plausibly favour
the published pruner.

\paragraph{The interaction estimator is unvalidated in magnitude.}
Pairwise and third-order terms come from a finite-difference estimator whose additive and
pairwise-only cases are exactly zero on synthetic set functions, but whose variance under the
one-sided degradation metric is not characterised. An earlier implementation carried a sign error; the
values in Figure~\ref{fig:interactions} use the corrected estimator and should be reproduced
independently before the magnitudes are relied upon.

\paragraph{The benchmark is circular.}
Its target circuit was found with the same simulator and the same objective, so recovery is a
consistency check rather than independent validation. Its objective threshold also yields one feasible
point rather than a frontier, which is why the evaluator must be extended before it can be used as
intended. Finally, masking in the released implementation does not compact the solver, so
retained-substrate cost does not imply runtime savings.

Adaptive sufficiency, plasticity, synapse-level search, whole-brain scaling, and sex comparison are
out of scope. We formalise adaptive sufficiency separately, where a retained parameter-norm trust
region stands in for a compensation budget, precisely to mark it as unsupported by the present
experiments.

\section{Conclusion}

We have formalised computational sufficiency as a certified, failure-constrained family of subsets
conditioned on a specified phenotype, separated task adequacy from phenotype fidelity, and introduced
measurements of non-decomposability, degeneracy, and functional role that single-circuit results
cannot express.

On the released walking simulator, the framework converts a published point estimate into a
certified statement: the published $148$-neuron circuit is sufficient with margin at the published
tolerance, no $146$- or $147$-neuron deletion subset of it was sufficient, its three interneurons are
individually necessary and not replaceable by their same-type bilateral partners, and a fourth
improves fidelity without being required. Adaptive complement splitting reaches that published size in
five of seven seeds, three of them recovering the published set exactly, in $66$--$97$ mask
evaluations, while the bundle hierarchy stops at $221$ under the same budget. The certified answer
also has structure a single circuit cannot show: the two interneurons $277$ and $617$ show a strong
saturating pairwise dependence, random pairs elsewhere in the same network have exactly zero
interaction, alternative $148$- and $149$-neuron solutions exist but only clear a looser tolerance,
and varying only the task changes which neurons are required. On the complete male CNS, the protected
interface alone fails every replicate, and adding ninety-three search-chosen neurons to its
$18{,}752$ yields a certified solution while random retention of eight times as many does worse.
Those experiments are complementary: the machinery certifies a known circuit when a known circuit
exists, and it finds a certified sparse augmentation at a scale where one-at-a-time methods cannot
run.

The same experiments draw the boundary of the method. On the
all-excitatory ventral-nerve-cord model the interface alone already reproduces most of the response,
so a near-empty circuit is trivially sufficient and the search is indistinguishable from random
removal. On controlled systems the hierarchy loses to greedy where exhaustive sweeps are affordable.
On the walking circuit the hierarchy loses to complement splitting where the necessary set is sparse,
yet the same splitting method is dominated by the hierarchy at whole-CNS scale, and the
sibling-batched variant of it recovers the published minimum on only one seed in four. No single
search strategy dominates, so the framework presents a toolkit, and no sufficiency frontier is
interpretable without the interface-only baseline next to it.

Sufficiency is therefore a set-valued property conditioned jointly on task and tolerance rather than a
single minimal circuit, and no one search strategy dominates across regimes, so a sufficiency claim is
only interpretable when reported with its interface-only baseline, its tolerance, and its search
budget. Every number in this paper is a property of a simulator under sampled parameters; the
framework is what we would defend, and the biology remains to be established.

\appendix

\section{Reproducibility}

The accompanying package is dependency-light and self-contained:
\begin{verbatim}
python -m pytest tests -q                                   # test suite
python -m approxbrain.experiments.cli --n 10 --budget 200 --seed 0
python hpc/run_frontier.py --budget 400 --duration 100       # VNC frontier
python hpc/download_malecns.py --outdir /data/male_cns      # needs token
python hpc/run_malecns_frontier.py --data /data/male_cns \
    --budget 1200 --duration 30 --tonic 0.55 --processes 24  # CNS frontier
python hpc/malecns_random_control.py --processes 20          # random control
python hpc/malecns_interactions.py --pairs 80 --processes 20
python hpc/pugliese_e0.py                                # reproduction check
python hpc/pugliese_search.py --outdir /data/pugliese/split \
    --budget 400 --split-parts 8 --methods split --tag split   # recovers 148
python hpc/pugliese_search.py --outdir /data/pugliese/rerun \
    --budget 400 --greedy-probe 32 --methods hcabes,greedy --tag dump
python hpc/pugliese_certify.py --outdir /data/pugliese/certify \
    --reps 32 --seed 7 --cohort-seed 7 --epsilon 0.05 --rho 0.2  # walking
python hpc/pugliese_certify.py --outdir /data/pugliese/certify_degeneracy \
    --reps 32 --seed 7 --cohort-seed 7 --battery degeneracy  # partner swaps
\end{verbatim}
All random draws are seed-controlled and propagated explicitly. The real-circuit experiments need
\texttt{manc\_dng100.npz} and \texttt{male\_cns/}, retrieved once through neuPrint and distributed
with the artefacts so the runs themselves need no token. The walking benchmark additionally needs a
checkout of the pinned upstream simulator (commit \texttt{faee4b0}), imported unchanged through
\texttt{hpc/pugliese\_adapter.py}; the certification battery for the recovered split mask adds
\texttt{--candidates-from} pointing at its \texttt{masks\_split\_split.jsonl}. Fork-based parallel
evaluation requires a POSIX host; on platforms without \texttt{fork} the same code runs sequentially
and produces identical results, which the test suite checks directly.

\bibliographystyle{unsrt}
\bibliography{main}

\end{document}